\documentclass[a4paper, 11pt]{article}
\usepackage[american]{babel}
\usepackage[T1]{fontenc}
\usepackage[utf8]{inputenc}
\usepackage{hyphenat}
\usepackage{longtable}
\usepackage{fullpage}
\usepackage{graphics}
\usepackage{natbib}
\usepackage{setspace}
\usepackage{tabularx}
\usepackage[table]{xcolor}
\usepackage{slashbox}
\usepackage{arydshln}
\usepackage{prettyref}
\usepackage{pifont}
\usepackage[strict]{changepage}
\usepackage{array}
\usepackage{booktabs}
\usepackage{multirow}
\usepackage{calc}
\usepackage{textcomp}
\usepackage{comment}
\usepackage{makecell}
\usepackage{sectsty}
\usepackage{epstopdf}
\usepackage{ccaption}
\captionnamefont{\bf}
\usepackage{bm}
\usepackage{pdfpages}
\usepackage{graphicx}
\usepackage[bitstream-charter]{mathdesign}
\usepackage{tikz}
\usepackage[figuresright]{rotating}
\usepackage{amsmath}
\numberwithin{equation}{section}
\usepackage[official]{eurosym}
\usepackage{pdflscape}
\usepackage{url}
\usepackage{wasysym}
\usepackage{bigdelim}
\usepackage{siunitx}
\usepackage[autolanguage]{numprint}
\usepackage{hhline}
\usepackage{csquotes}
\usepackage[amsmath]{ntheorem}
\usepackage{subcaption}

\usepackage[bottom]{footmisc}
\usepackage{fancyhdr}   
\usepackage{enumitem}
\usepackage{chngcntr}   
\usepackage[colorlinks, linkcolor = black, citecolor = black, filecolor = black, urlcolor = black]{hyperref}            

\usepackage{pdfpages}

\newcommand{\GRAPHSCITE}{X/Graphs/}

\newcommand{\GRAPHSPAPER}{X/Graphs/}

\newcommand{\LabelUSCit}{US citations}
\newcommand{\LabelEPCit}{EP citations}
\newcommand{\LabelKogan}{\citet{kogan2017technological} (USD)}
\newcommand{\LabelPatVal}{PatVal (EUR)}
\newcommand{\LabelUSClaims}{US claim length}
\newcommand{\LabelEPClaims}{EP claim length}

\newcommand\WoSTotal{\numprint{42962463}}

\newcommand\WoSTotalM{\numprint{43}}

\newcommand\WoSTotalNoSoc{\numprint{35874824}}

\newcommand\WoShasSNPLFiveY{\numprint{1627872}}

\newcommand\WoShasSNPLFiveYNoSoc{\numprint{1597426}}

\newcommand\WoShasSNPLFiveYNoAsc{\numprint{1465312}}

\newcommand\WoShasSNPL{\numprint{2248563}}

\newcommand\WoShasSNPLM{\numprint{2.2}}

\newcommand\WoShasSNPLNoSoc{\numprint{2203035}}

\newcommand\WoShasSNPLNoAsc{\numprint{2079713}}

\newcommand\ScopusSNPLUnifiedGrant{\numprint{113340}}

\newcommand\ScopusSNPLUnifiedGrantUnique{\numprint{49254}}

\newcommand\PatFamFrontierFiveTotal{\numprint{3816176}}

\newcommand\PatFamTotal{\numprint{4767844}}

\newcommand\PatFamTotalM{\numprint{4.8}}

\newcommand\PatFamSNPLTotal{\numprint{952932}}

\newcommand\PatFamSNPLTotalM{\numprint{1}}

\newcommand\PatFamEPTotal{\numprint{1960772}}

\newcommand\PatFamUSTotal{\numprint{4442742}}

\newcommand\PatFamSNPLEPTotal{\numprint{490848}}

\newcommand\PatFamSNPLUSTotal{\numprint{921929}}

\newcommand\PatFamManySNPLShare{\numprint{64.2}}

\newcommand\PatFamAvgSNPL{\numprint{7.4}}

\newcommand\PatFamXiTotal{\numprint{1029987}}

\newcommand\PatFamXiSNPLTotal{\numprint{230389}}

\newcommand\PatFamPatValTotal{\numprint{11061}}

\newcommand\PatFamPatValSNPLTotal{\numprint{2579}}

\newcommand\SNPLEPAll{\numprint{1009481}}

\newcommand\SNPLEPAllUnique{\numprint{575637}}

\newcommand\SNPLUSAll{\numprint{6177977}}

\newcommand\SNPLUSAllUnique{\numprint{2017694}}

\newcommand\SNPLUnifiedAll{\numprint{6962239}}

\newcommand\SNPLUnifiedAllM{\numprint{7.0}}

\newcommand\SNPLUnifiedAllUnique{\numprint{2229658}}

\newcommand\PrecisionEPO{\numprint{0.99}}

\newcommand\RecallEPO{\numprint{0.96}}

\newcommand\PrecisionUSPTO{\numprint{0.99}}

\newcommand\RecallUSPTO{\numprint{0.92}}

\newcommand\PrecisionWIPO{\numprint{0.99}}

\newcommand\RecallWIPO{\numprint{0.97}}

\newcommand\MeanResidUSCitValZero{\numprint{5.122}}
\newcommand\MeanResidUSCitValOne{\numprint{4.906}}
\newcommand\MeanResidUSCitValTen{\numprint{10.107}}
\newcommand\MeanResidUSCitStdErrZero{\numprint{0.004}}
\newcommand\MeanResidUSCitStdErrOne{\numprint{0.022}}
\newcommand\MeanResidUSCitStdErrTen{\numprint{0.057}}
\newcommand\MeanResidUSCitCountZero{\numprint{  3466695}}
\newcommand\MeanResidUSCitCountOne{\numprint{    84839}}
\newcommand\MeanResidUSCitCountTen{\numprint{    85262}}

\newcommand\MeanResidEPCitValZero{\numprint{0.947}}
\newcommand\MeanResidEPCitValOne{\numprint{0.746}}
\newcommand\MeanResidEPCitValTen{\numprint{2.068}}
\newcommand\MeanResidEPCitStdErrZero{\numprint{0.001}}
\newcommand\MeanResidEPCitStdErrOne{\numprint{0.012}}
\newcommand\MeanResidEPCitStdErrTen{\numprint{0.016}}
\newcommand\MeanResidEPCitCountZero{\numprint{  3466695}}
\newcommand\MeanResidEPCitCountOne{\numprint{    84839}}
\newcommand\MeanResidEPCitCountTen{\numprint{    85262}}

\newcommand\MeanResidKoganValZero{\numprint{13.322}}
\newcommand\MeanResidKoganValOne{\numprint{12.411}}
\newcommand\MeanResidKoganValTen{\numprint{16.796}}
\newcommand\MeanResidKoganStdErrZero{\numprint{0.044}}
\newcommand\MeanResidKoganStdErrOne{\numprint{0.622}}
\newcommand\MeanResidKoganStdErrTen{\numprint{0.475}}
\newcommand\MeanResidKoganCountZero{\numprint{   699752}}
\newcommand\MeanResidKoganCountOne{\numprint{     8899}}
\newcommand\MeanResidKoganCountTen{\numprint{    13665}}

\newcommand\MeanResidPatValValZero{\numprint{11.859}}
\newcommand\MeanResidPatValValOne{\numprint{8.401}}
\newcommand\MeanResidPatValValTen{\numprint{23.923}}
\newcommand\MeanResidPatValStdErrZero{\numprint{0.449}}
\newcommand\MeanResidPatValStdErrOne{\numprint{3.209}}
\newcommand\MeanResidPatValStdErrTen{\numprint{4.891}}
\newcommand\MeanResidPatValCountZero{\numprint{     8482}}
\newcommand\MeanResidPatValCountOne{\numprint{      351}}
\newcommand\MeanResidPatValCountTen{\numprint{      232}}

\newcommand\MeanResidUSClaimsValZero{\numprint{185.547}}
\newcommand\MeanResidUSClaimsValOne{\numprint{179.163}}
\newcommand\MeanResidUSClaimsValTen{\numprint{177.672}}
\newcommand\MeanResidUSClaimsStdErrZero{\numprint{0.082}}
\newcommand\MeanResidUSClaimsStdErrOne{\numprint{0.455}}
\newcommand\MeanResidUSClaimsStdErrTen{\numprint{0.491}}
\newcommand\MeanResidUSClaimsCountZero{\numprint{  1952410}}
\newcommand\MeanResidUSClaimsCountOne{\numprint{    66313}}
\newcommand\MeanResidUSClaimsCountTen{\numprint{    70738}}

\newcommand\MeanResidEPClaimsValZero{\numprint{143.926}}
\newcommand\MeanResidEPClaimsValOne{\numprint{140.834}}
\newcommand\MeanResidEPClaimsValTen{\numprint{128.968}}
\newcommand\MeanResidEPClaimsStdErrZero{\numprint{0.084}}
\newcommand\MeanResidEPClaimsStdErrOne{\numprint{0.333}}
\newcommand\MeanResidEPClaimsStdErrTen{\numprint{0.454}}
\newcommand\MeanResidEPClaimsCountZero{\numprint{  1157166}}
\newcommand\MeanResidEPClaimsCountOne{\numprint{    42849}}
\newcommand\MeanResidEPClaimsCountTen{\numprint{    30109}}

\begin{document}
\title{\textbf{Science Quality and the Value of Inventions}\thanks{The authors are listed in reverse alphabetical order. We gratefully acknowledge support from the Max Planck Society and the German Science Foundation (DFG, CRC TRR 190). We thank Wolfgang Knaus, Birgit Palzenberger and Frank Sander for their support in matching WoS data to patent NPL references and Jian Wang for helpful comments on the computation of citation metrics. Earlier results from this research were presented at the Web of Science Data Workshop at EPFL and the 2018 Annual Conference of EPIP (European Policy for Intellectual Property). The Web of Science data are available via Thomson Reuters. The patent data are available via the PATSTAT dataset provided by the European Patent Office.
Supplementary materials: attached below. Figs. \ref{supp:fig:precision:recall} to \ref{supp:fig:interdisciplinarity}. Tables \ref{supp:tab:desc:counts} to \ref{reg:pat:sci:cit:lag}. 
}
} 

\vspace{-3.5cm}
\author{
\begin{minipage}{0.25\textwidth}
    \begin{center}
    \vspace{0.1cm}
    Felix Poege$\,^{ab}$
    \vspace{0.6cm}
    \end{center}
\end{minipage}
\begin{minipage}{0.25\textwidth}
    \begin{center}
    \vspace{0.1cm}
    Dietmar Harhoff$\,^{acd}$\thanks{Corresponding author: Max Planck Institute for Innovation and Competition, Marstallplatz 1, D-80539 Munich. Mail: \href{mailto:dietmar.harhoff@ip.mpg.de}{dietmar.harhoff@ip.mpg.de}.}
    \vspace{0.6cm}
    \end{center}
\end{minipage}
\begin{minipage}{0.25\textwidth}
    \begin{center}
    \vspace{0.1cm}
    Fabian Gaessler$\,^{ae}$
    \vspace{0.6cm}
    \end{center}
\end{minipage}
\begin{minipage}{0.25\textwidth}
    \begin{center}
    \vspace{0.1cm}
    Stefano Baruffaldi$\,^{af}$
    \vspace{0.6cm}
    \end{center}
\end{minipage}
\\[2ex]
\renewcommand{\parskip}{-5pt}
\footnotesize $^{a}$  \emph{Max Planck Institute for Innovation and Competition, Munich}\\
\footnotesize $^{b}$  \emph{Institute for the Study of Labor (IZA), Bonn}\\
\footnotesize $^{c}$  \emph{Ludwig Maximilian University (LMU), Munich}\\
\footnotesize $^{d}$  \emph{Centre for Economic Policy Research (CEPR), London}\\
\footnotesize $^{e}$  \emph{Technical University of Munich, Munich}\\
\footnotesize $^{f}$  \emph{University of Bath, Bath}\\
[3ex]}

\date{\today\\[1cm]
}
\maketitle
\vspace{5mm}
\thispagestyle{empty}

\begin{center}
\subsubsection*{ABSTRACT}
\end{center}
{\small
\it
Despite decades of research, the relationship between the quality of science and the value of inventions has remained unclear. We present the result of a large-scale matching exercise between \PatFamTotalM{} million patent families and \WoSTotalM{} million publication records. We find a strong positive relationship between quality of scientific contributions referenced in patents and the value of the respective inventions. We rank patents by the quality of the science they are linked to. Strikingly, high-rank patents are twice as valuable as low-rank patents, which in turn are about as valuable as patents without direct science link. We show this core result for various science quality and patent value measures. The effect of science quality on patent value remains relevant even when science is linked indirectly through other patents. Our findings imply that what is considered “excellent” within the science sector also leads to outstanding outcomes in the technological or commercial realm.
}
\vspace{10mm}
\textbf{KEYWORDS:} science, citations, non-patent literature, patents, patent value, patent families.
\vspace{2mm}\\
\textbf{JEL Classification:} L20, O31, O33, O34

\onehalfspacing
\newpage
\clearpage
\setcounter{page}{1}

\section{Introduction}

The relationship between science and technology has been subject to intense discussions for centuries. Science was largely funded via patronage during the Renaissance, and a separation of public funding for fundamental research and private, industrial funding for applied research and commercial innovation efforts only emerged in the 19th century \citep{Scotchmer2004, Mokyr2002}. Since the aftermath of World War II, policy-makers have relied on the notion that science helps to generate knowledge and information which will ultimately contribute to the emergence of new technical and organizational capabilities, improvements in the quality of life and economic growth \citep{bush1945science}. Vannevar Bush’s vision of a publicly funded science system that feeds into privately organized innovation channels became the blueprint for most of the Western national systems of science funding, R\&D and innovation. This notion has recently come under scrutiny again as voters increasingly demand evidence on the benefits of science spending. For policy-makers and scientists alike, it is tantamount to improve the understanding of the impact of science on technical progress and innovation.

The most pertinent form of output delivered by the science sector are publications, which are known to vary widely in quality. While some scientific publications will reach and inspire large numbers of researchers, others are never read or referenced. Measures of science quality, such as citation counts or impact factors, are used to make this heterogeneity visible and have become increasingly important in the governance of the science sector. Science governance and science funding seek to promote excellent over more mediocre science output by allocating resources to those researchers and institutions from which outstanding results can be expected.

But it has been argued that this logic does not take tangible results from technology transfer and commercialization into account. Science is inward-looking according to these voices. This raises the question to what extent science output that is considered ``excellent'' within the science sector can lead to outstanding outcomes in the technological or commercial realm. This paper seeks to contribute new insights toward the understanding of this nexus.

We provide evidence that the quality of scientific publications -- as commonly assessed in science via citations -- is a strong predictor of their relevance for and impact on technology development as documented in patents. We document two main results. First, publications with high scientific quality are vastly more likely to be cited in patent documents, and cited at a higher rate. Second, among patents directly building on science, the value of patents increases monotonically with science quality. These results hold across scientific disciplines, technology areas and time. 

\section{Data}
\label{sec:data}

Our analysis starts from the universe of scientific publications in Web of Science (WoS) from the year 1980 onwards, corresponding to approximately \WoSTotalM{}  million scientific publications. In terms of patents, we consider a sample of more than \PatFamTotalM{} million patent families, comprising all patent families from the database DOCDB with at least one grant publication at the European Patent Office (EPO) or the United States Patent and Trademark Office (USPTO), with first filing date between 1985 and 2012, included. Subsequently, our unit of analysis is the patent family, to which we also interchangeably refer as patents. The patents protect inventions in developed countries with in total more than one billion inhabitants.

Patents reference various types of documents which relate to the protected material either by determining novelty (prior art) or by explaining the content of the underlying invention. These documents include foremost other patents, but frequently also non-patent literature (NPL). \citep{callaert2006traces} A subset of the latter are references to scientific articles, which we dub Scientific Non-Patent Literature (SNPL).

To link patents to publications, we leverage a highly precise and comprehensive match of NPL references in patents with scientific publications in WoS.\footnote{The match is based on a methodology documented in detail in \citet{knaus_parma_2018} and summarized in the supplementary material.} The NPL references in patents that were successfully link to scientific publications comprise our set of \textit{S}NPL references. Around \PatFamSNPLTotalM{} million patents were linked to at least one scientific publication via a total of about \SNPLUnifiedAllM{} million SNPL references. Out of all scientific publications, about \WoShasSNPLM{} million figure in this list of SNPL references.

In our core set of analyses, we rely on established measures of scientific quality and patent value. The quality of scientific publications is measured by the number of citations from other scientific publications over a period of three years from publication. We define a patent’s SNPL science quality as the quality of the patent’s SNPL references. A patent can reference zero, one or several scientific articles, in the same way as a scientific article can be referenced by zero, one or many patents. Figure \ref{fig:setting} illustrates this setup. When more than one SNPL reference is present, we consider by default only the publication with the highest quality. Patent value is measured by the number of forward patent citations over a period of five years from the patent's first filing date. We use citations by US patents as our first measure of patent value. Our results are robust to alternative choices. We replace citations as science quality measure by the journal impact factor. We replace our aggregation method of the quality of multiple SNPL references with several other options. We replace US patent citations as value measure by a host of alternatives. \footnote{The supplementary material provides further detailed information on data sources, discusses the use of citations as indicators of relatedness between technology and science and elaborates on alternative measures of patent value as well as scientific quality that we use for robustness analyses.}

\section{Results}
\label{sec:results}

\begin{figure}[htb]
\begin{subfigure}{\textwidth}
    \centering
    \includegraphics[width=.9\linewidth]{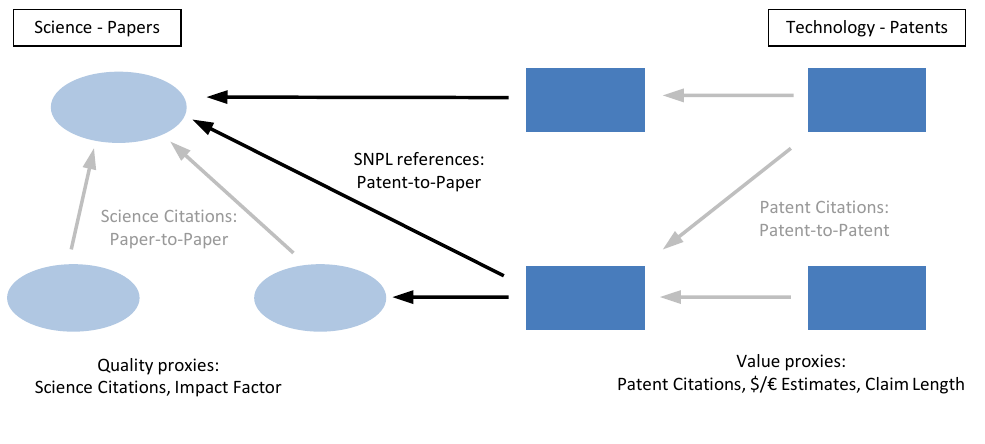}
    \caption{Setting: The domains of science (left), technology (right) and patent-paper references \label{fig:setting}}
\end{subfigure} \\
\begin{subfigure}{.5\textwidth}
    \centering
    \includegraphics[width=\linewidth]{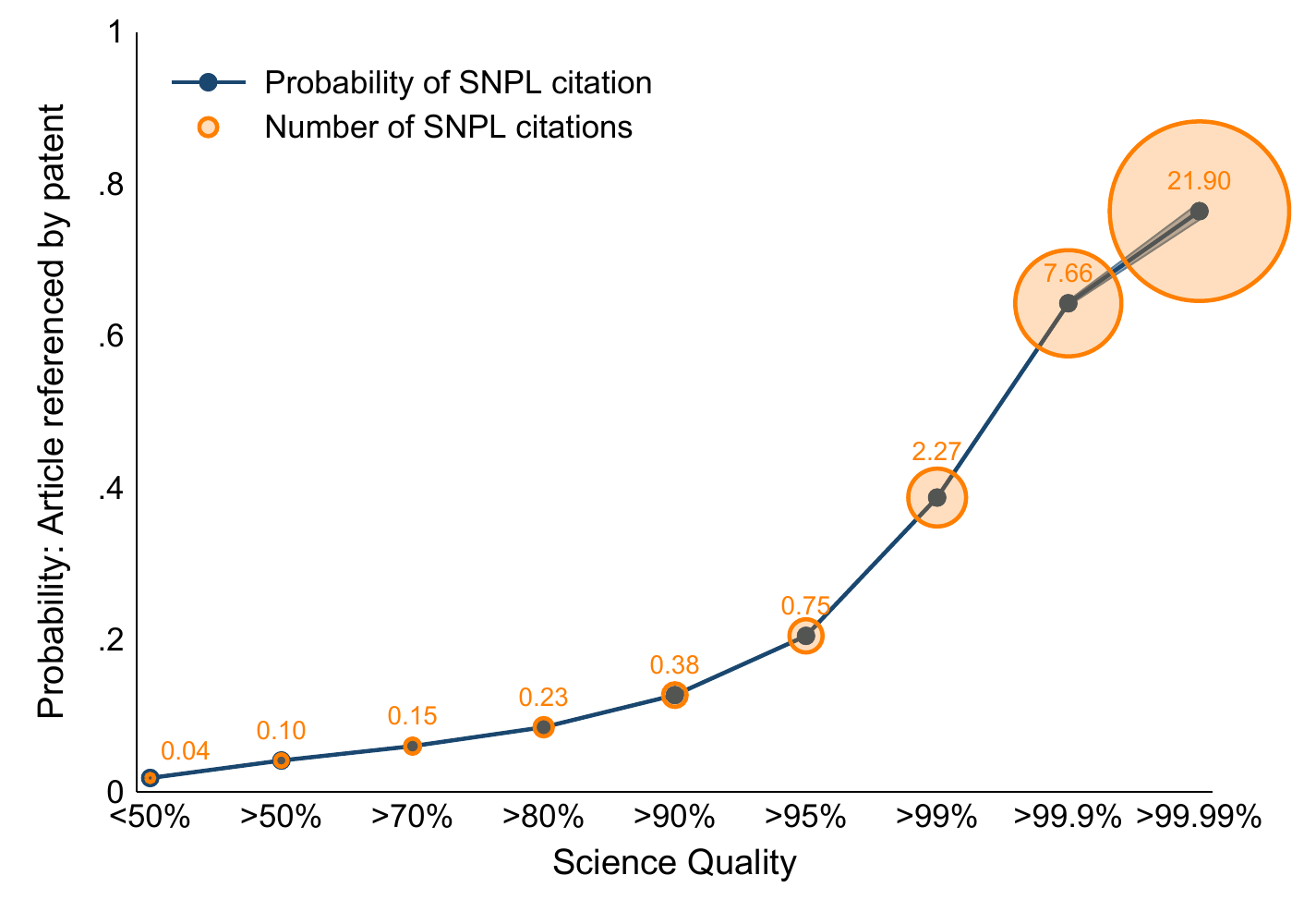}
    \caption{SNPL references by science quality} \label{fig:paper:selection}
\end{subfigure}%
\begin{subfigure}{.5\textwidth}
    \centering
    \includegraphics[width=\linewidth]{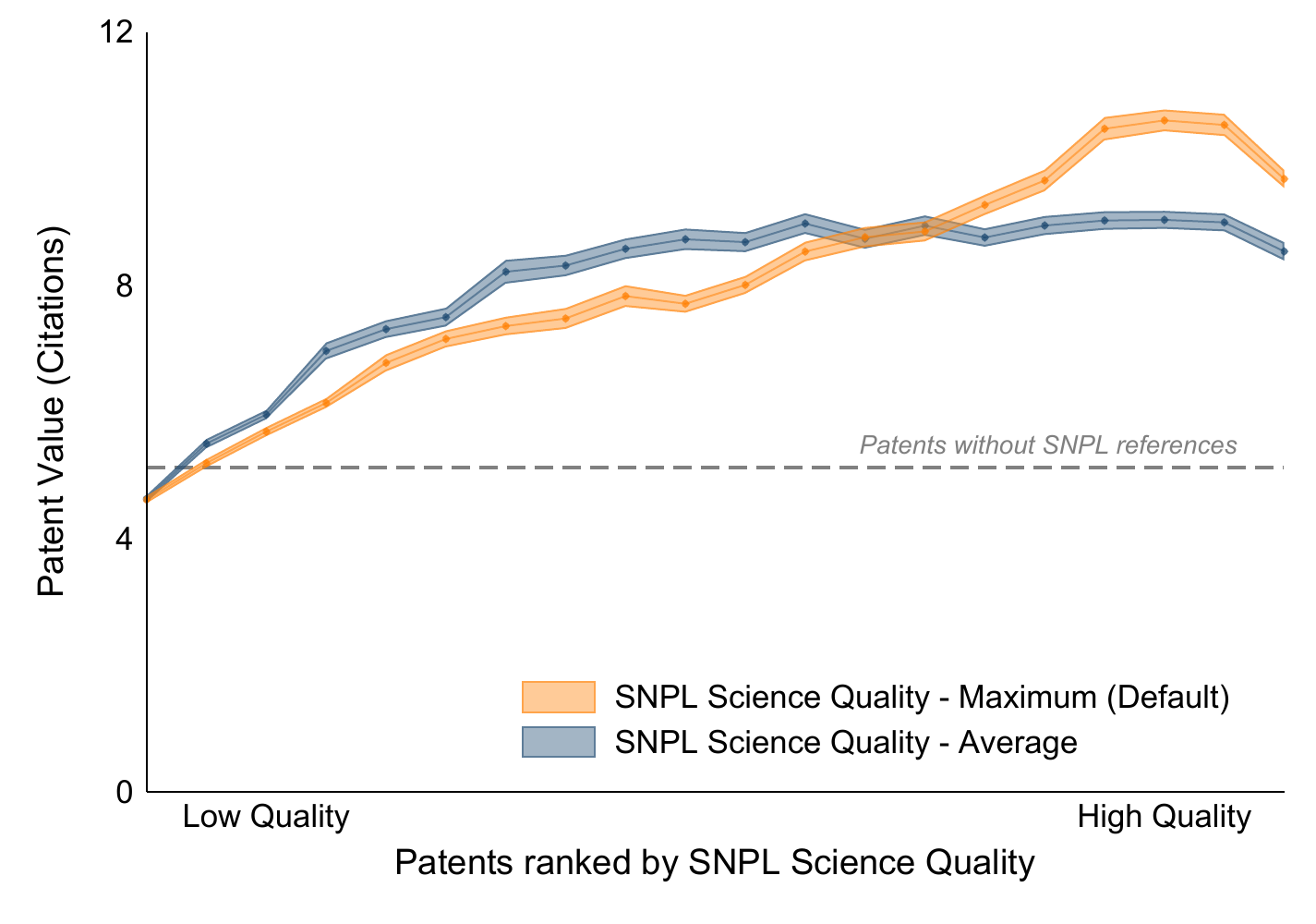}
    \caption{Patent value by SNPL science quality \label{fig:paper:percentiles}}
\end{subfigure}
\caption{Setting and main results}
\legend{\footnotesize{\textbf{Notes:} Science quality is the 3-year citation count from other scientific publications. b) The patent count is not conditional on appearing as a SNPL reference. Blue shaded areas show 95\% confidence intervals around the mean. N = \WoSTotal{}.
c) SNPL science quality is the quality of publications referenced by a patent. When there are multiple patent-paper references, we by default use the highest-quality reference (orange). In comparison, we use the average quality (blue). Patent value is measured as the 5 year count of patent forward citations by US patents. Patent value and science quality are residualized using technology field $\times$ first filing year FEs. Shaded areas show 95\% confidence intervals around the respective means. N = \PatFamTotal{} patents (\PatFamSNPLTotal{} with SNPL references).}}
\end{figure}

As a first-order question, we explore the selection of scientific publications into the patent realm, i.e., the relationship between science quality and the likelihood that a scientific publication is referenced in a patent \citep{hicks2000research}. We look at the probability and intensity of referencing, i.e., if any and how many patented inventions refer to a given scientific contribution. We present results for publications below the median (all receiving 0 science citations), for publications between the median and the 70th percentile, and at the percentiles 80, 90, 95, 99 (top 1 percent), 99.9 (top 1 permille), and 99.99 (top permyriad) of scientific quality. Figure \ref{fig:paper:selection} presents these results; the line plots the share of scientific publications appearing as SNPL reference in at least one patent, and the size of the circles indicates the average number of times they appear as SNPL references.

We find a remarkably strong positive selection of scientific publications of high scientific quality into SNPL references. Below the median, scientific publications are almost never SNPL references. This number increases up to 40\% at the top 1\% of publications by scientific quality. A staggering majority of publications at the top 1 permille (>60\%) and beyond the top 1 permyriad (80\%) are referenced in patents. The average number of times they appear as SNPL references in distinct patent families is 7.7 and 21.9, respectively. 
We emphasize that these results are not due to a feedback from important patents to citations of the underlying science. By restricting our measure for scientific citations to the first three years after publication, we have effectively excluded this bias.

We move on to our main analysis and investigate the extent to which SNPL science quality is a predictor of patent value. The main figures account for level differences across technology fields and over time. We estimate econometric models that absorb variation across these dimensions with pair-level fixed-effect (FE) controls and graphically present the resulting residual values. In effect, we transform deviations from the technology field and year specific mean to deviations from the overall mean. This ensures that structural differences across technological fields and over time do not drive the results.
The relationships discussed are backed up by econometric models that allow quantifying their average magnitude, assessing their statistical significance, and controlling for a full set of confounding factors.

The relationship between SNPL science quality and patent value is depicted in figure \ref{fig:paper:percentiles}. We plot the average patent value across the distribution of SNPL science quality. As a first measure of patent value we use the number of patent citations from US patents. We later on consider alternatives. As a benchmark level, the figure shows the average value of patents without any SNPL reference (dashed line). We contrast two possible aggregation methods of SNPL science quality. When a patent references multiple scientific articles, we in a first variant use highest-quality reference as our measure (orange). Here, we juxtapose a second variant, where we consider the average quality of all references. Clearly, top science matters much more, considering scientific material beyond the best one dilutes the science quality-technology value relationship. In the supplementary material, we show that this extends to using other aggregation methods which focus on the top of the quality distribution. Consequently, we continue by only considering the highest-quality SNPL reference.

Previous studies have encountered a higher value of patents with SNPL references or references to other technical literature, in limited samples or specific fields \citep{branstetter2005exploring,harhoff2003citations}. We are able to confirm this finding, on a large scale, in our data: the value of patents with SNPL references is higher, or equal than the value of patents without SNPL references, for any level of SNPL science quality except the very bottom.

Notably, SNPL science quality fully explains the difference in value between patents with and without SNPL references. Patent value increases rapidly, and almost monotonically, for a higher level of SNPL science quality. Patents with SNPL references at the bottom of the SNPL science quality distribution are on average as valuable as patents without SNPL references. Compared to this group, patents at the top of the SNPL science quality distribution receive more than twice as many forward patent citations. This core result suggests that scientific activities of high quality lead to the development of highly valuable technologies.

\begin{figure}[htb]
\begin{subfigure}{.5\textwidth}
    \centering
    \includegraphics[width=\linewidth]{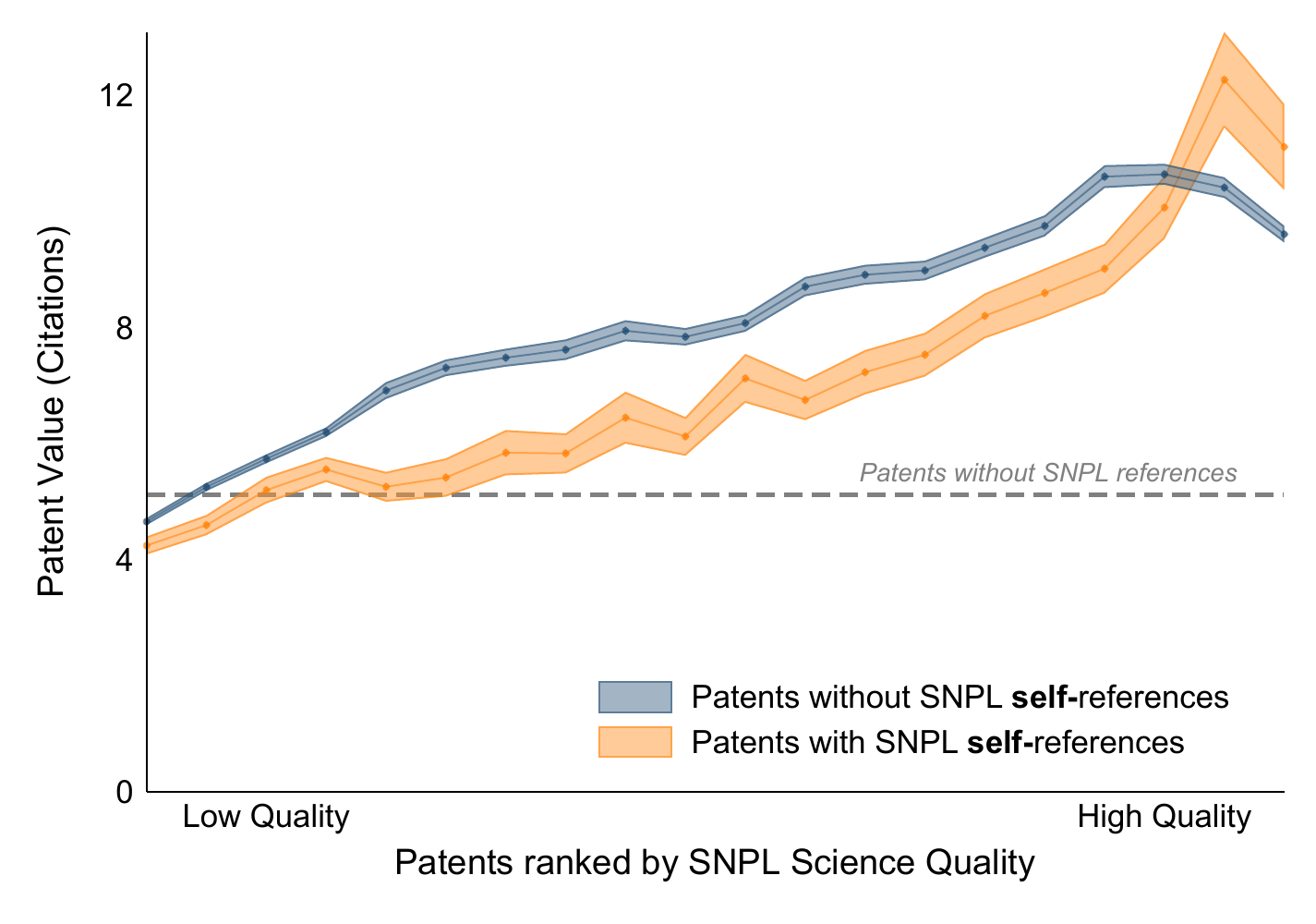}
    \caption{SNPL self-references \label{fig:paper:self}}
\end{subfigure}
\begin{subfigure}{.5\textwidth}
    \centering
    \addtocounter{subfigure}{1} 
    \includegraphics[width=\linewidth]{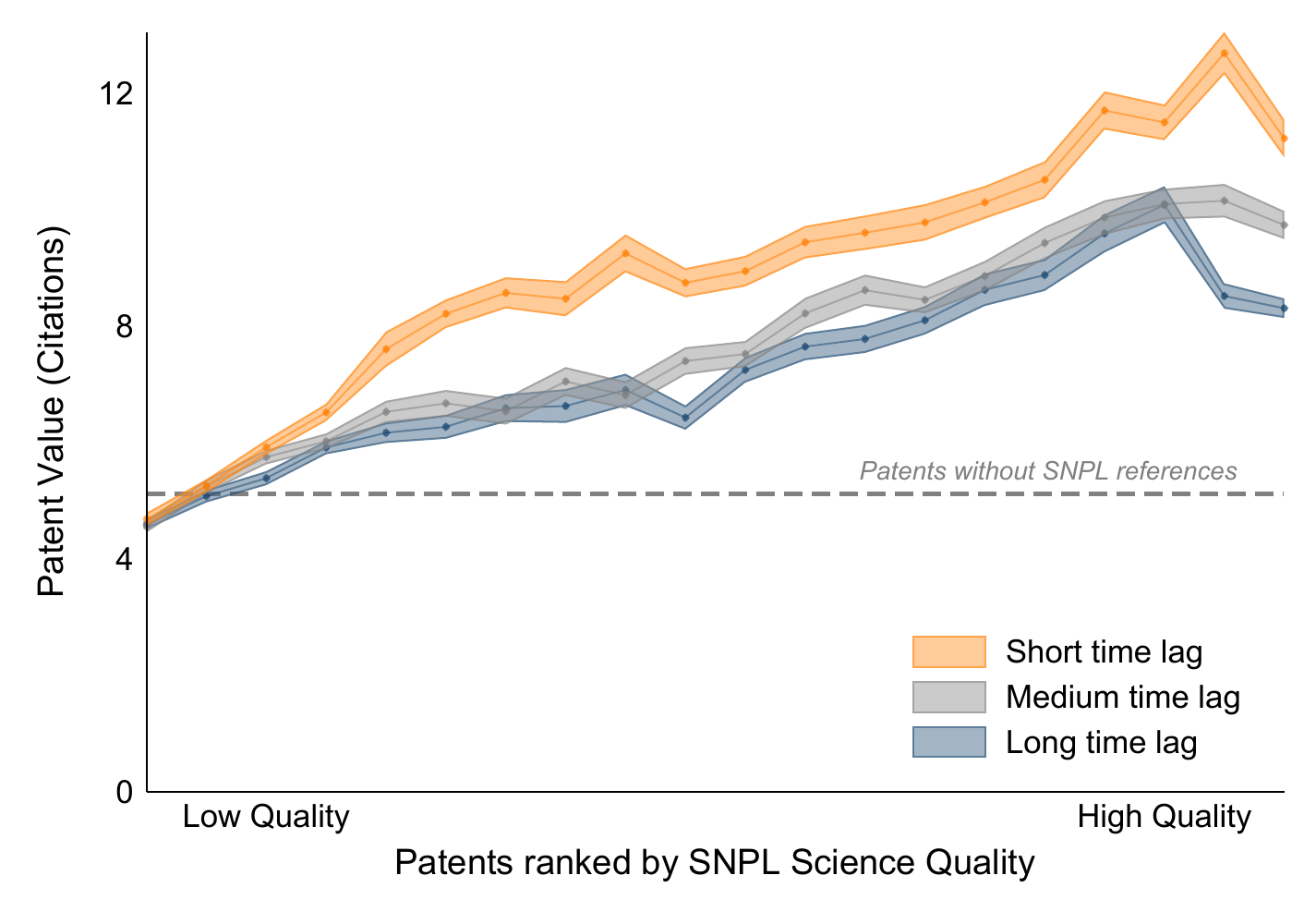}
    \caption{Patent value by SNPL science quality and time} \label{fig:paper:lag}
\end{subfigure}
\vspace{-4pt}
\begin{center}
    \begin{subfigure}{.7\textwidth}
        \centering
        \addtocounter{subfigure}{-2} 
        \includegraphics[width=\linewidth]{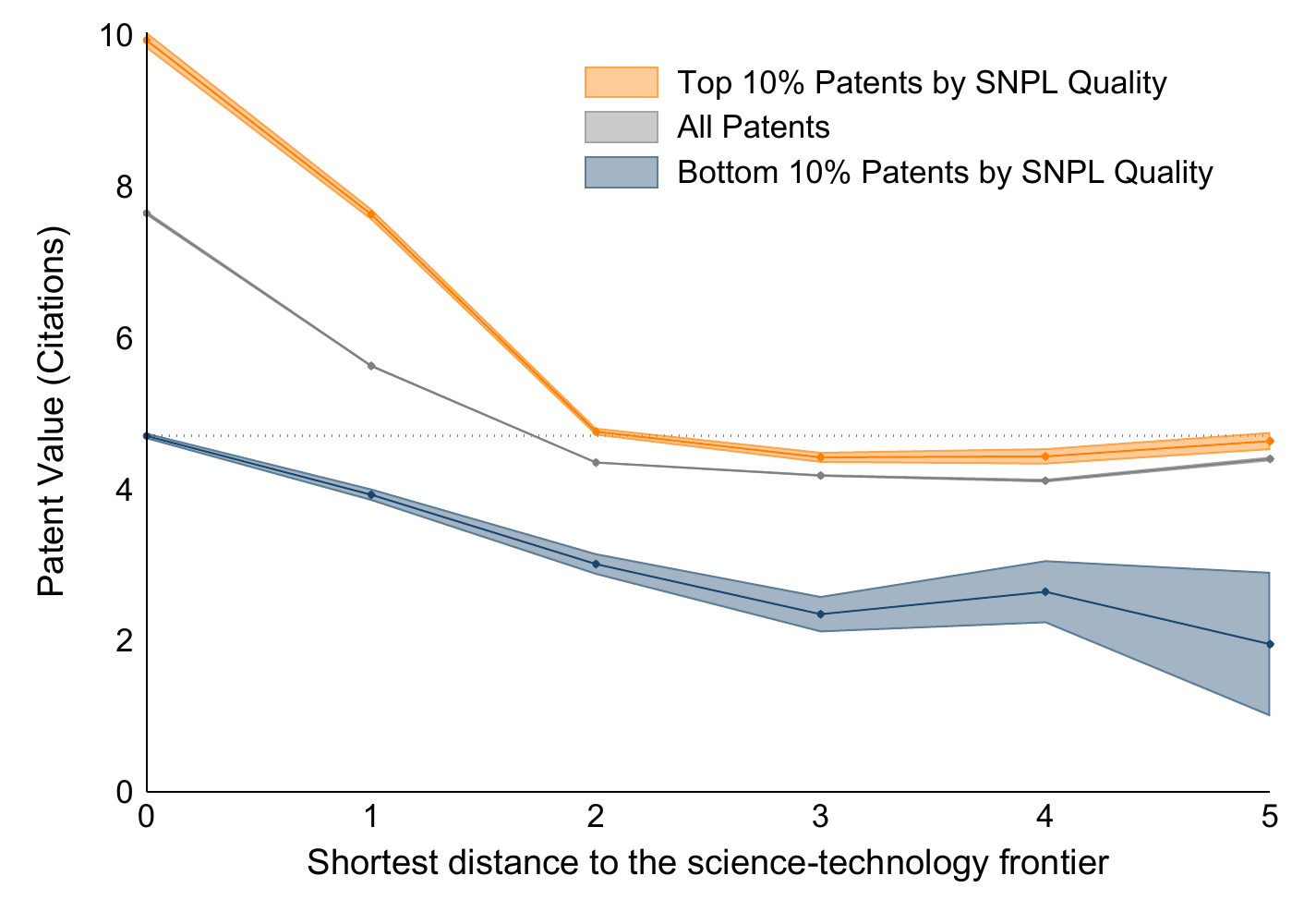}
        \caption{Patent value by distance to the scientific frontier and SNPL science quality \label{fig:paper:distance}}
    \end{subfigure}
\end{center}
\vspace{-20pt}
\caption{Additional results}
\vspace{-8pt}
\legend{\footnotesize{\textbf{Notes:} SNPL science quality is the maximum 3 year citation count across scientific publications appearing as SNPL references in a patent. Patent value is measured as the 5 year count of patent forward citations by US patents. Patent value and science quality are residualized using technology field $\times$ first filing year FEs. Shaded areas show 95\% confidence intervals around the respective means.

a) SNPL self-references of the highest-quality SNPL reference are considered. N = \PatFamTotal{} patents (\PatFamSNPLTotal{} with SNPL references).

b) The distance to the science frontier (x-axis) is measured as the shortest path to a patent with SNPL references in the patent references network. For patents not at the science frontier, SNPL science quality is the maximum SNPL science quality in patents at the frontier to which they are linked. N = \PatFamFrontierFiveTotal{}

c) Time-distance is measured as the lag between the first filing year of the patent and the publication year of the scientific publication in SNPL references with the the highest science quality.  N = \PatFamTotal{} patents (\PatFamSNPLTotal{} with SNPL references).
}}
\end{figure}

Possibly, high quality research and technology development are undertaken by the same individuals or organizations, which may drive the result. Companies, startups, inventors and academic scientists can perform scientific activities that may lead directly to both scientific and technological outcomes \citep{gittelman2003does}. Therefore, we complement this finding exploring how our results vary considering separately SNPL self-references. 
Figure \ref{fig:paper:self} describes the corresponding results. The line in orange indicates the patent value of patents with SNPL self-references. The line in blue describes the value of patents excluding SNPL self-references. The latter presents close to identical results to those obtained in figure \ref{fig:paper:percentiles}. Note that part of the SNPL science quality distribution, with the exception of the very top, patent value is higher if patents with SNPL self-references are excluded. In fact, the share of SNPL self-references is roughly similar and, if anything, tends to decrease for higher levels of SNPL science quality. Overall this is supportive of the idea that high-quality science is linked to high-value technology also, and especially, beyond the organizational boundaries within which it is developed.

Our analysis so far has focused on patents at the frontier with science, i.e., linked directly to a scientific publication via an SNPL reference. To generalize our findings, we also consider patents connected to scientific publications indirectly via references to other patents. Patents for which the shortest path in the citation network is longer are said to be more distant from the science-technology frontier. Recent studies have used this concept of distance between science and technology, and demonstrate that the value of patents monotonically decreases for higher distances to the science frontier \citep{ahmadpoor2017dual}. In Figure \ref{fig:paper:distance} we consider this dimension and describe the value of patents at different levels of distance from the science-technology frontier. We distinguish patents linked (directly or indirectly) to SNPL references at the top 10\% and bottom 10\% of quality. We also report the average value of all patents, at different distances. Patents linked to more than one patent with SNPL references at the same distance are assigned to the patent with the highest-quality SNPL reference.

We find that the correlation between patent value and SNPL science quality largely propagates to patents at higher distances from the science-technology frontier. The increase in patent value for a change from the average patent to patents at the top 10\% (patents citing other patents with SNPL references to scientific publications of high quality) is approximately equal to the increase in patent value for a step closer to the frontier. For instance, patents at one step of distance from the top 10\% have the same value than the average patent at the frontier with science. Patents at any distance from the top 10\% always have higher values than patents at the bottom 10\%. The difference persists also at a high distance from the frontier, approximately constant and equal to about a 3 times higher value. Regression results in the supplementary material confirm that the positive correlation between SNPL science quality and patent value starts fading only after a degree of distance higher than 6. We can conclude that science of high quality spurs technological progress of high value far beyond the science-technology frontier.

In figure \ref{fig:paper:lag} we also consider time as a related dimension to distance from science. Time is measured as the lag between the first filing year of a patent family and the publication year of the highest-quality SNPL reference. We study how patent value varies along the SNPL science quality distribution and for different levels of time lag. Interestingly, shorter time lags are always associated with higher patent value. The correlation with SNPL science quality remains strongly positive for all levels of time-distance, but is stronger for patents with short time-distance. As a consequence, at high levels of SNPL science quality, patent value is high, on average, but increases also sharply for shorter time lags. Conversely, at low levels of SNPL science quality the marginal effect of time-distance is small.

So far, we have measured patent value with US forward patent citations. However, the results are robust across a broad set of alternative measures of patent value. First, we consider the count of citations from the EPO. Second, we adopt two indicators of monetary value, available for a subsample of patents. We use estimates from Kogan et al. \citep{kogan2017technological}, who propose a measure based on abnormal stock market returns at the patent's grant event as a proxy for its private value. We further obtain inventor survey-based value estimates of patented inventions from the PatVal survey \citep{giuri2007inventors}. These two measures are only available for a limited sample of patents of about 899k and 11k, respectively. Third, we measure patent scope by the length of the text of the first independent claim. This relies on evidence showing that longer descriptions of the claimed invention implies more narrow legal protection and, therefore, a lower patent value \citep{kuhn2019scope}. We consider separately, and when available, the length of the first independent claim in the patent grant publication at the USPTO or the EPO. Table \ref{tab:value} reports descriptive statistics based on the average of all these alternative patent value indicators for patents without SNPL references, and for patents in the top 10\% and bottom 10\% of SNPL science quality. We replicate regression results for all these alternative measures of patent value in the supplementary material.

\begin{table}[ht]
    \setlength{\tabcolsep}{12pt}
    \caption{SNPL science quality and alternative measures of patent value\label{tab:value}}
    \centering
    \setlength\extrarowheight{-1pt}
    \renewcommand{\arraystretch}{1.05}
    \begin{tabular}{@{}l*{3}{S}@{}} \toprule
        &   {\small{No SNPL}}&{\small{Bottom 10\%}}&  {\small{Top 10\%}}\\
        \midrule
 \small{\LabelUSCit} \\
 \, \footnotesize{\textbf{Mean}} & \footnotesize{\textbf{\MeanResidUSCitValZero}} & \footnotesize{\textbf{\MeanResidUSCitValOne}} & \footnotesize{\textbf{\MeanResidUSCitValTen}} \\
 \, \footnotesize{Standard Error} & \footnotesize{(\MeanResidUSCitStdErrZero)} & \footnotesize{(\MeanResidUSCitStdErrOne)} & \footnotesize{(\MeanResidUSCitStdErrTen)} \\
 \, \footnotesize{N} & \footnotesize{{\MeanResidUSCitCountZero}} & \footnotesize{{\MeanResidUSCitCountOne}} & \footnotesize{{\MeanResidUSCitCountTen}} \\
 \hdashline[0.5pt/5pt]
 
 \small{\LabelEPCit} \\
 \, \footnotesize{\textbf{Mean}} & \footnotesize{\textbf{\MeanResidEPCitValZero}} & \footnotesize{\textbf{\MeanResidEPCitValOne}} & \footnotesize{\textbf{\MeanResidEPCitValTen}} \\
 \, \footnotesize{Standard Error} & \footnotesize{(\MeanResidEPCitStdErrZero)} & \footnotesize{(\MeanResidEPCitStdErrOne)} & \footnotesize{(\MeanResidEPCitStdErrTen)} \\
 \, \footnotesize{N} & \footnotesize{{\MeanResidEPCitCountZero}} & \footnotesize{{\MeanResidEPCitCountOne}} & \footnotesize{{\MeanResidEPCitCountTen}} \\
 \hdashline[0.5pt/5pt]

 \small{\LabelKogan} \\
 \, \footnotesize{\textbf{Mean}} & \footnotesize{\textbf{\MeanResidKoganValZero}} & \footnotesize{\textbf{\MeanResidKoganValOne}} & \footnotesize{\textbf{\MeanResidKoganValTen}} \\
 \, \footnotesize{Standard Error} & \footnotesize{(\MeanResidKoganStdErrZero)} & \footnotesize{(\MeanResidKoganStdErrOne)} & \footnotesize{(\MeanResidKoganStdErrTen)} \\
 \, \footnotesize{N} & \footnotesize{{\MeanResidKoganCountZero}} & \footnotesize{{\MeanResidKoganCountOne}} & \footnotesize{{\MeanResidKoganCountTen}} \\
 \hdashline[0.5pt/5pt]

 \small{\LabelPatVal} \\
 \, \footnotesize{\textbf{Mean}} & \footnotesize{\textbf{\MeanResidPatValValZero}} & \footnotesize{\textbf{\MeanResidPatValValOne}} & \footnotesize{\textbf{\MeanResidPatValValTen}} \\
 \, \footnotesize{Standard Error} & \footnotesize{(\MeanResidPatValStdErrZero)} & \footnotesize{(\MeanResidPatValStdErrOne)} & \footnotesize{(\MeanResidPatValStdErrTen)} \\ 
 \, \footnotesize{N} & \footnotesize{{\MeanResidPatValCountZero}} & \footnotesize{{\MeanResidPatValCountOne}} & \footnotesize{{\MeanResidPatValCountTen}} \\
 \hdashline[0.5pt/5pt]

 \small{\LabelUSClaims} \\
 \, \footnotesize{\textbf{Mean}} & \footnotesize{\textbf{\MeanResidUSClaimsValZero}} & \footnotesize{\textbf{\MeanResidUSClaimsValOne}} & \footnotesize{\textbf{\MeanResidUSClaimsValTen}} \\
 \, \footnotesize{Standard Error} & \footnotesize{(\MeanResidUSClaimsStdErrZero)} & \footnotesize{(\MeanResidUSClaimsStdErrOne)} & \footnotesize{(\MeanResidUSClaimsStdErrTen)} \\
 \, \footnotesize{N} & \footnotesize{{\MeanResidUSClaimsCountZero}} & \footnotesize{{\MeanResidUSClaimsCountOne}} & \footnotesize{{\MeanResidUSClaimsCountTen}} \\
 \hdashline[0.5pt/5pt]

 \small{\LabelEPClaims} \\
 \, \footnotesize{\textbf{Mean}} & \footnotesize{\textbf{\MeanResidEPClaimsValZero}} & \footnotesize{\textbf{\MeanResidEPClaimsValOne}} & \footnotesize{\textbf{\MeanResidEPClaimsValTen}} \\
 \, \footnotesize{Standard Error} & \footnotesize{(\MeanResidEPClaimsStdErrZero)} & \footnotesize{(\MeanResidEPClaimsStdErrOne)} & \footnotesize{(\MeanResidEPClaimsStdErrTen)} \\
 \, \footnotesize{N} & \footnotesize{{\MeanResidEPClaimsCountZero}} & \footnotesize{{\MeanResidEPClaimsCountOne}} & \footnotesize{{\MeanResidEPClaimsCountTen}} \\
 \bottomrule
        \end{tabular}
    \vspace{4pt}    
    \legend{\footnotesize{\textbf{Notes:} The table presents descriptive statistics for all considered measures of patent value. It reports average values for patents without SNPL references, with SNPL references in the bottom 10\% and in the top 10\% of science quality. Patent value and science quality are residualized using technology field $\times$ year FEs. Elasticities from corresponding regression analysis are available in the supplementary material.}}
\end{table}
\clearpage

\section{Conclusions}
\label{sec:conclusion}

The quality of scientific contributions is often measured in terms of their impact within the scientific community. Yet, scientists also need to gauge and acknowledge of their contributions for society and future technical and social advancements. The fact that science quality is practically defined within the realm of science itself, contributes to a perception of science as being an independent upstream activity, at times detached from technological progress, with an indirect and delayed impact on society at best.

To the contrary, our study suggests that such an interpretation of the relationship between science quality and technology would largely be a misconception. We show that excellent science is directly linked to inventions of particularly high value. More specifically, our findings demonstrate that there is a robust and strong relationship between the scientific quality of a publication referenced in a patent and the patent’s impact and commercial value.

Our results are descriptive, and the exact causes of the strong correlation will have to be analyzed in future work. At this point, it seems most reasonable to presume that industrial users of scientific insights scan the science sector for novel results and employ the ones that are most promising for applications in their industrial fields. We doubt that they do so merely on the basis of science citation counts or impact measures. Rather, we expect that they apply their own complex logic and assessments, and that they may even avoid using the classical metrics of the science sector altogether. Commercial investments are unlikely to be made on the premise the citation-measured interest in the scientific community was sufficiently high. Hence, the high correlation between quality measures used in the science sector and those used in the commercial (patent) realm are fortuitous. They are highly unlikely to reflect a spurious selection result.

Leaving aside the exact causal links, our results provide intriguing evidence for the governance system of science, e.g. at universities and public research organizations, as well as for funding agencies and science policy-makers. The current system steers researchers to strive for success measured in terms of citations and impact. According to our findings, the outcomes of such a system are well-aligned with later stages of technology development and translation of science results. Our study does not provide evidence on the optimality of the alignment. However, it clearly contradicts the notion that the application of scientific criteria in science funding decisions would lead researchers to engage in exercises that are of little value to society at large. Quite to the contrary, science quality (as measured by scientists) is a strong predictor of applicability and practical value of the technologies developed as the fruits of scientific endeavor. Paradoxically, when making commercial investment decisions, taking academic measures such as citation counts or impact factors into account may not be a bad idea.

\clearpage

\begin{center}
{\huge Supplementary material for:\\[2ex]
    \textbf{Science Quality and the Value of Inventions}
}
\end{center}

\thispagestyle{empty}

\vfill

\subsection*{This document includes:}

{\small
\begin{itemize}
\item Sections \ref{suppl:sec:materials_and_methods} to \ref{suppl:sec:suppl_graphs_tables}
\vspace{-5pt}
\begin{itemize}
    \item Data
    \item Methods (incl. literature review)
    \item Regression analyses
    \item Supplementary graphs and tables
\end{itemize}
\item Figs \ref{supp:fig:precision:recall} to \ref{supp:fig:interdisciplinarity}
\item Tables \ref{supp:tab:desc:counts} to \ref{reg:pat:sci:cit:lag}
\item References
\end{itemize}
}

\onehalfspacing
\newpage
\clearpage
\setcounter{page}{12}

\setcounter{section}{0}
\setcounter{figure}{0}
\setcounter{table}{0}
\renewcommand{\thefigure}{S-F\arabic{figure}}
\renewcommand{\thetable}{S-T\arabic{table}}
\renewcommand{\thesection}{S-\arabic{section}}
\renewcommand{\thesubsection}{\thesection.\arabic{subsection}}

\section{Data}
\label{suppl:sec:materials_and_methods}

\begin{figure}[htb]
    \centering
    \caption{Sample descriptives over time \label{supp:fig:desc:time}}
        \includegraphics[width=0.7\linewidth]{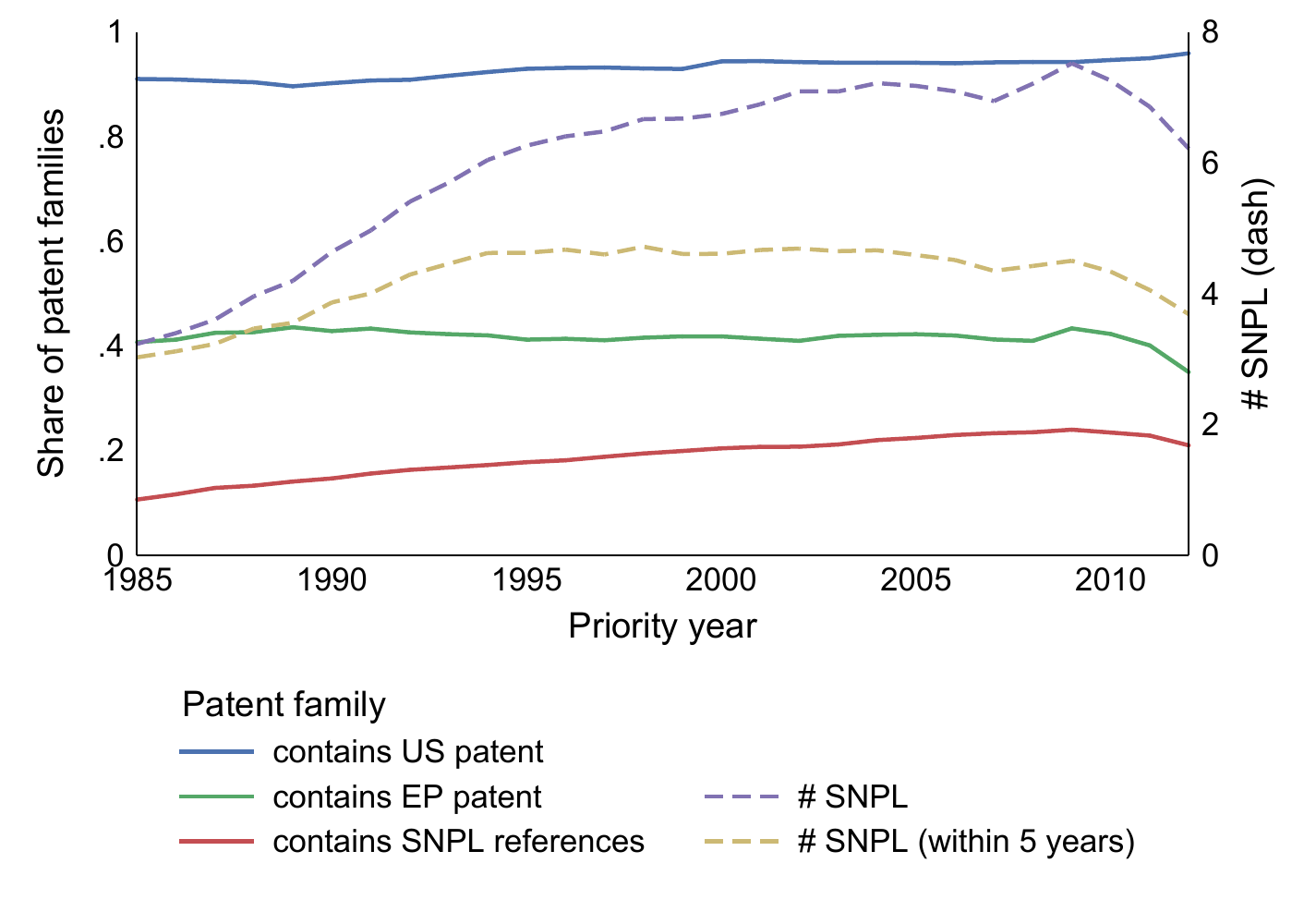}
    \legend{\footnotesize{\textbf{Notes:} Shows the composition of the estimation sample. The estimation sample contains US and EP patents from patent families with at least one member patent granted at USPTO or EPO.
}}
\end{figure}

In the following we briefly introduce the scientific literature and patent data. Table \ref{supp:tab:desc:counts} provides details on the structure of the merged dataset. Figure \ref{supp:fig:desc:time} shows descriptive statistics over time on the samples of patents and SNPL references.

\subsubsection*{Scientific literature data}
\label{suppl:sec:scientific_literature_data}

Scientific literature data comes from \WoSTotalM\,million scientific publications, corresponding to all research articles indexed in the Thomson Reuters Web of Science (WoS) database that were published between 1980 and 2016. WoS is the largest bibliographic database of scientific literature and provides all main information for each scientific publication, including authors, affiliations, research field and citations.\footnote{More extensive information on the WoS is available at \href{www.webofknowledge.com}{www.webofknowledge.com}.}

\subsubsection*{Patent data}
\label{suppl:sec:patent_data}

The main source of patent data in our study is the database DOCDB, a database maintained and updated on a weekly basis by the European Patent Office (EPO).\footnote{More extensive information on DOCDB is available at \href{www.epo.org/searching-for-patents/data/bulk-data-sets/docdb}{www.epo.org/searching-for-patents/data/bulk-data-sets/docdb}} It includes records from more than 90 patent offices. We base our study on a sample of more than \PatFamTotalM{} million patent families in DOCDB, comprising all patent families with at least one grant publication at the European Patent Office (EPO) or the United States Patent and Trademark Office (USPTO), with first filing date between 1985 and 2012, included. We include references generated during the search and examination phase of patents filed at the EPO, USPTO or the World Intellectual Property Organization (WIPO). Note that at the WIPO, there is no grant procedure and WIPO examinations are typically conducted by the EPO.

DOCDB contains all information digitally available on these patents. An advantage with respect to non-patent literature (NPL) citations data, as compared to other databases, is the availability of enriched xml text comprising separate fields for title, authors, year, journals title, pages, volume and number. This allows matching this information separately with bibliographic scientific literature information, substantially improving the quality of the match (see section \ref{suppl:sec:linking_npl}). 

Whenever we refer to technology field, we use the classification of IPC patent codes in the 34 technology fields provided by WIPO.\footnote{WIPO Classification: \href{https://www.wipo.int/edocs/mdocs/classifications/en/ipc_ce_41/ipc_ce_41_5-annex1.pdf}{https://www.wipo.int/edocs/mdocs/classifications/en/ipc\_ce\_41/ipc\_ce\_41\_5-annex1.pdf}}

\begin{table}[ht]
    \setlength{\tabcolsep}{8pt}
    \caption{Structure of the dataset\label{supp:tab:desc:counts}}
    \centering
    \small
    \renewcommand{\arraystretch}{1.2}
    \begin{tabular}{@{}p{7cm}rrr@{}} \toprule
        Scientific publications (1980-2012) &   \multicolumn{1}{c}{ \qquad Total} & \multicolumn{1}{c}{Excluding} & \multicolumn{1}{c}{Excluding}\\
        &   \multicolumn{1}{c}{} & \multicolumn{1}{c}{social/humanities} & \multicolumn{1}{c}{self-references}\\        
        \midrule
        Scientific publications &        \WoSTotal & \WoSTotalNoSoc & \\
        Scientific publications in SNPL references &        \WoShasSNPL & \WoShasSNPLNoSoc & \WoShasSNPLNoAsc \\
        Scientific publications in SNPL references (within five years) &        \WoShasSNPLFiveY & \WoShasSNPLFiveYNoSoc & \WoShasSNPLFiveYNoAsc \\
     
        \midrule
        Patent families (1985-2012) &   \multicolumn{1}{c}{ \qquad Total} & \multicolumn{1}{c}{ \qquad EPO} & \multicolumn{1}{c}{ \qquad USPTO}\\
        \midrule
        Patent family - SNPL reference combinations  &        \SNPLUnifiedAll           & \SNPLEPAll            & \SNPLUSAll \\
        Unique SNPL references                       &        \SNPLUnifiedAllUnique     & \SNPLEPAllUnique      & \SNPLUSAllUnique \\
        Patent families                                 &        \PatFamTotal              & \PatFamEPTotal        & \PatFamUSTotal \\
        Patent families with SNPL references            &        \PatFamSNPLTotal          & \PatFamSNPLEPTotal    & \PatFamSNPLUSTotal \\
        
        \bottomrule
    \end{tabular}
    \legend{\footnotesize{\textbf{Notes:} Observation counts in the dataset. Discrepancies originate from the different views on the data. The first part of the table also considers SNPL citations from the 1980-1984 range, whereas the second part does not.}}
\end{table}

\subsubsection*{Data transformation}

Whenever we use logarithmic transformations on variables with natural zero values (e.g. citation counts), we use a log($x+1$) transformation.
When unifying patent attributes at the patent family level, several decisions have to be taken. For technology fields, we use the modal technology field of member patents. In case of ties, we use the numerically lowest field. When no field classification is available, we drop the patent family.
When multiple patent value estimates from \citet{kogan2017technological} or PatVal \citep{giuri2007inventors} are available, we use the highest one.
Some variables with extreme values are winsorized. Backward reference counts, the number of times a patent references to other patents, are winsorized at the 95th percentile. The same is applied for the number of inventors and SNPL references. Lengths of the first independent claim are winsorized at the 1st and 99th percentile.
When assigning scientific fields to scientific publications, in case of multiple fields, we retain the scientific field whose codes are first in the alphabet.
We restrict our sample to SNPL citations where the publication year of the scientific article was at or before the first filing year of the patent family.

\section{Methods}
\label{suppl:sec:methods}

\subsection{Linking Scientific and Patent Literature Data}
\label{suppl:sec:linking_npl}

``Science'' usually refers to the creation and organization of knowledge, often in the form of testable hypotheses and predictions regarding natural phenomena. In a stark simplification, academic scientists (who are mostly employed in the public sector) live in a world governed by the quest for making pioneering contributions to knowledge, hence striving for novelty of insight and for a better understanding of fundamental issues \citep{merton1968matthew, merton1973sociology}. According to this view, scientists also follow norms of disclosing newly generated knowledge and information in scientific publications. The societal or private benefit from applications is considered less important, but also hard to assess directly. In principle, the science could thus be decoupled from the economic pursuit of wealth and monetary gain.  

Conversely, ``technology'' refers to the realm of the artificial and to artifacts which may have, or may have not, been constructed with the help of scientific insights. Technology is defined in the OECD Frascati Manual as the collection of techniques, skills, methods, and processes used when producing goods and services. Applications of new insights are largely brought about by engineers  \citep{Allen1977}. Engineers (who mostly work in the private sector) are governed by rules and incentives that are very different from those guiding the behavior of scientists. They seek to contribute new technologies, use secrecy to protect the market positions of their employers and are involved in strategic considerations of market rivalry. Engineers thus turn knowledge into marketable products which then generate monetary returns for owners. This by now classical view of the relationship between science and technology is described, inter alia, by \citet{Allen1977} and \citet{Brooks1994}.  

Initially understood as two distinct and independent realms, science is now viewed to directly facilitate the application of new knowledge \citep{stokes2011pasteur}, and that science and technology may follow a process of co-evolution \citep{murray2002innovation}. Science has also been described as a kind of map used in the process of devising new technologies \citep{fleming2004science}. This new view acknowledges  that the realms of science and commercial technology development overlap and that their relationship is not necessarily a linear one. While universities mostly generate knowledge, they also file patent applications and license intellectual property. And corporate entities mostly seek to commercialize new products and services, but also engage in basic research not immediately tied to product development and in publication of research results.

\subsubsection*{SNPL references as a measure of knowledge input}

We use non-patent literature references to scientific publications (SNPL) as an indicator of relatedness of a technology, as described in a patent, to scientific contributions, as reported in scientific publications. Numerous studies have proposed patent citations as an indicator of knowledge flows \citep{jaffe1986technological,jaffe1989real}. While some authors have raised concerns on the validity of this approach for general patent citations \citep{thompson2005patent,alcacer2006patent}, SNPL references have been consistently found to be more related to actual knowledge flows than other types of references \citep{roach2013lens}. In the context of our study, it is not necessary to interpret SNPL references as a direct indicator of knowledge flows: we assume more broadly that a cited scientific paper contains relevant information for the understanding and the development of a technology.

\subsubsection*{SNPL matching methodology}

The dataset we adopt to link patents to cited scientific publications is a full match of DOCDB patent data with bibliographic information included in WoS. The matching process is documented in detail in \citet{knaus_parma_2018}. Here we present a brief overview.

The matching consists of three steps, target selection, search and quality control. In the target selection step, cleaning steps are undertaken to exclude NPL strings which are no scientific articles or are outside of the available WoS data. For the remaining entries, a search engine was employed to look up NPL full-text strings in a full-text index of the complete WoS or Scopus content. The search engine returns a ranked list of match candidates. During the quality control stage, the topmost candidate is examined and the match quality is judged according to a field-based scoring. Only high-quality matches are considered valid matches for the final dataset.

The matching procedure is applied on a first set of roughly 37 million NPL references. 27 million (71.8\%) entries were selected as a potential target and linked to WoS entries. However, not all of these constitute a valid match after taking the quality of the match into account. The quality of a match is judged by six quality indicators (year, volume, page(s), first author, journal title, article title). Each of these indicators equals one if the information from the matched scientific article can be found in the non-patent literature citation string. The quality score is the sum of the indicators and ranges from zero to six. 

\begin{figure}[htb]
    \centering
    \caption{Precision-recall tradeoff in the WoS match\label{supp:fig:precision:recall}}
        \includegraphics[width=0.7\linewidth]{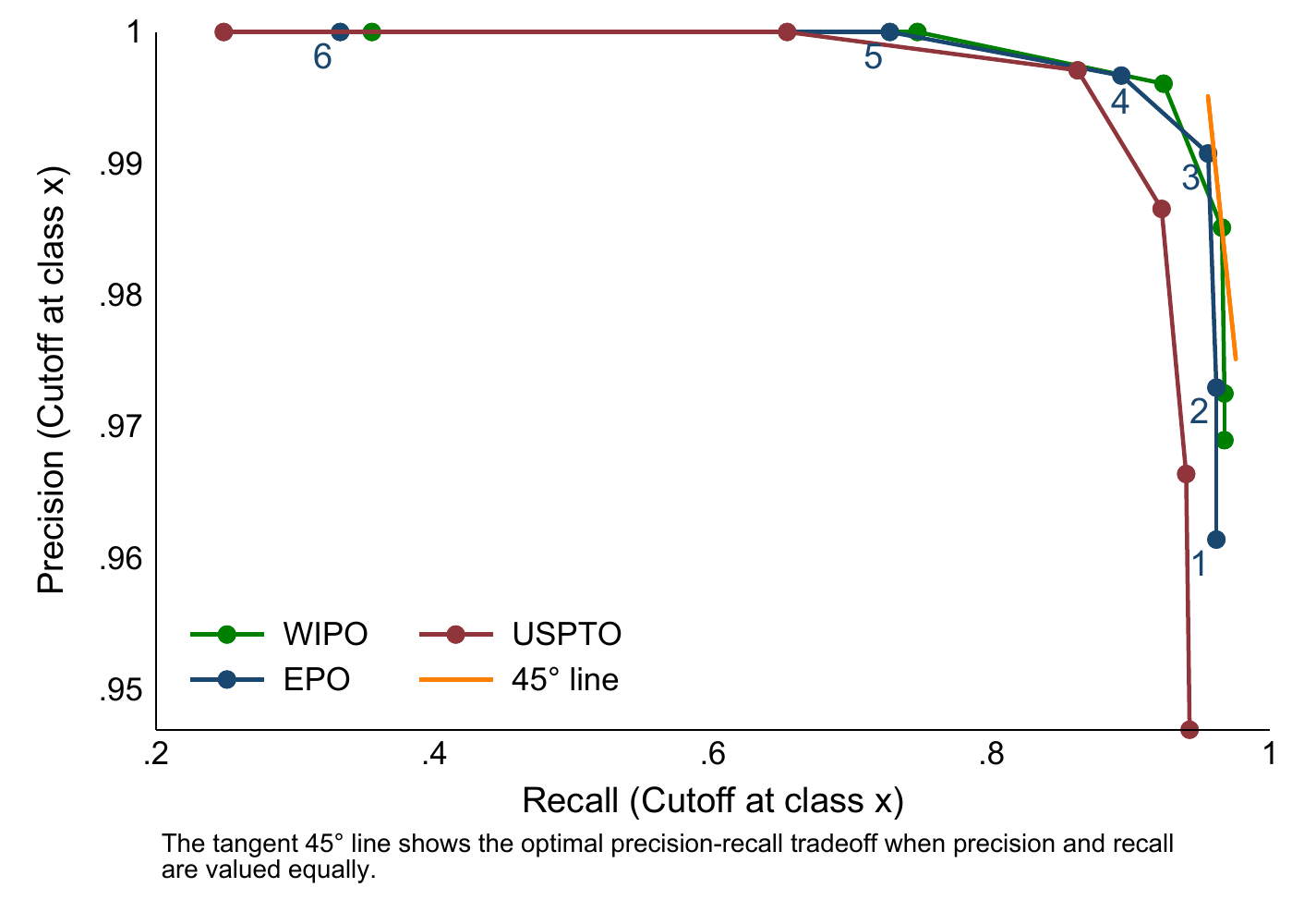}
    \legend{\footnotesize{\textbf{Notes:} Based on manually evaluating 1000 matches for each patent office, as reported in \citet{knaus_parma_2018}. The 45°-line shows the set of points which is of the same quality if precision and recall are weighted equally. This corresponds to a F1-score. The point where this line is intersected is the optimal point, here shown for the WIPO validation. Precision is the share of SNPL reference matched correctly among the matched SNPL references. Recall is the ratio between correct matches identified and all SNPL references that were or should have been matched.}}
\end{figure}

To validate the matching quality, random subsamples of 1,000 NPL references each were drawn. An NPL string is considered a valid target if it could be found in the WoS using a manual search. Figure \ref{supp:fig:precision:recall} plots precision and recall, where precision is computed as the share of correct matches out of all matches delivered by the algorithm. Recall is the share of all targets which could be recovered successfully. The graph reveals that when accepting a quality score of three and higher as high-quality matches, precision scores of \PrecisionEPO\, and recall scores of \RecallEPO\, (EPO) and \RecallUSPTO\, (USPTO) can be achieved.\footnote{With a quality cutoff at four, the precision increases even further, but recall suffers to a greater extent so that the quality cutoff at three is preferred when putting equal weight on precision and recall.} Table \ref{supp:tab:desc:precision:recall} shows the final quality achieved.

We, therefore, restricted the sample to matches of quality equal to or higher than three. Out of the 27 million references retained as valid targets, 13 million (47.1\%) satisfied this quality requirement. Our units of analysis are DOCDB patent families which typically include multiple references.

While the precision and recall scores are high, they only refer to what could have been matched -- the content of the Web of Science. Clearly, not all scientific publications that can be referenced in patents are covered in this database. We assess the extent of this issue and consider the subset of NPL references which could not be matched to WoS. We attempt a match to an alternative publication database, Scopus, which has a larger coverage. This exercise generates \ScopusSNPLUnifiedGrant\, additional SNPL links to \ScopusSNPLUnifiedGrantUnique\, Scopus items for publication years 1996-2016. Given that this is less than 2\% of the total, for simplicity, we disregard these links in our analysis.

Our final sample contains \PatFamSNPLTotal\, DOCDB patent families with at least one grant publication and at least one matched SNPL reference at any of the patent offices considered here.

This compares well with previous datasets, and in general, constitute a larger number of observations than previously identified in existing studies. \citet{ahmadpoor2017dual} use patent data exclusively at the USPTO between 1976 and 2015 where 759,000 patents were found to be directly linked to at least one scientific publication in WoS via an NPL reference. \citet{jefferson2018mapping} starts with 11.8 million scientific publications published between 1980 and 2015, of which roughly 1.2 million are cited in 690,000 patent \textit{families} (1.1 million patents). \citet{marx_reliance_2019} link US patents from 1926-2018 to scientific papers from 1800-2018, identifying approximately 15.7 million citation links between 1.4 million patents to 2.9 million papers.
In comparison, our dataset links \PatFamSNPLTotal{} patent \textit{families} from 1985-2017 to \SNPLUnifiedAllUnique{} distinct scientific articles in the time range of 1980-2016.

\begin{table}[htb]
    \centering
    \setlength{\tabcolsep}{8pt}
    \caption{Match quality\label{supp:tab:desc:precision:recall}}
    \renewcommand{\arraystretch}{1.2}
    \begin{tabular}{@{}lSS@{}} \toprule
        \multicolumn{1}{l}{Office} & \multicolumn{1}{c}{Precision} & \multicolumn{1}{r}{Recall}\\
        \midrule
        EPO   &        \PrecisionEPO   & \RecallEPO \\
        USPTO &        \PrecisionUSPTO & \RecallUSPTO \\
        WIPO &        \PrecisionWIPO   & \RecallWIPO \\
        \bottomrule
    \end{tabular}
    \vspace{4pt}
    \legend{\footnotesize{\textbf{Notes:} Based on a manual validation exercise of 1000 NPL references per office, as reported in \citet{knaus_parma_2018}. Precision is the share of NPL reference matches that was correct. Recall is, when considering all NPL references that could have been matched, the share that were matched correctly.}}
\end{table}

\subsubsection*{SNPL self-references}

We single out SNPL references to scientific publications where at least one author also figures among the inventors of the patent and where one affiliation of the SNPL references overlap with the list of applicants in the patent. We refer to these categories as SNPL inventor self-references and applicant self-references, respectively. This type of SNPL references reveals links between patents and scientific publications originating from either the same organization or from the same individuals, or both.  The first analyses rely on the full sample of SNPL references. We present results separately for these categories and excluding them in a later stage.

We consider SNPL inventor self-references those that refer to scientific publications where at least one author has the same name of an inventor on the patent. We consider as SNPL applicant self-references those that refer to scientific publications where at least one affiliation overlaps with the list of applicants in the patent. To match applicants with affiliations we use a list of manually disambiguated organizations (academic institutions and firms) derived from the combination of multiple sources: the Global Research Identifier Database (GRID), the ORBIS database and the EU Scoreboards database. We merge separately applicants in patents and affiliations in scientific publications to these lists using a probabilistic matching algorithm based on training data.

We consider an applicant and an affiliation to be the same when they match to the same entity in the list. Note that the two categories of self-references may overlap.\footnote{Figure \ref{supp:fig:paper:self:desc} presents related descriptive statistics.}

 \subsubsection*{Related literature on SNPL references}
\label{suppl:sec:related_literature}

We briefly summarize the literature that has so far discussed the characteristics of SNPL references and their relationship with patent value. 

\citet{hicks2000research} look at all scientific articles published between 1993 and 1995 in journals indexed in the Science Citation Index (SCI) with at least one US author. They find that about 6,600 of these publications were cited in 1997 US-invented patents. The probability of a publication being cited as SNPL depends not only on the publication’s research field, but also on its scientific impact. If a publication belongs to the top 1\% most highly cited publications, it is about nine times more likely to be cited by a US patent than a randomly chosen US publication. In similar vein, \citet{popp2017science} finds in green energy technology fields that scientific articles that are cited frequently by other articles are also more likely to be cited by patents.

\citet{breschi2010tracing} analyze all patent applications to the European Patent Office (EPO) registered in the period 1990 to 2003 within three technology fields (lasers, semiconductors and biotechnology) and find about 44,000 patents with altogether 18,000 SNPL references. SNPL references are more frequent in biotech and lasers than in semiconductors, presumably due to the larger distance between the semiconductor technology field and science.

\citet{harhoff2003citations} are among the first to analyze the relationship between the value of such patents and the scientific impact of the underlying scientific contributions. They document a positive relationship between patent value and the number of NPL references. The relationship is particularly strong in the technical area of chemicals and pharmaceuticals. Several other authors have explored the role of NPL references as potential determinants of patent value. \citet{branstetter2005exploring} use a random sample of 30,000 US patents from 1983-86, of which about 4,300 include SNPL. Those patents that cite scientific articles are of significantly higher quality (more claims and forward-citations) than those that do not. \citet{sorenson2004science} link about 17,300 patents from 1990 with about 16,700 non-patent references. Here, patents that cite non-patent literature receive more citations and are cited more quickly than other patent. They argue this positive relationship between forward-citations and science intensity of a given patent is due to knowledge diffusion through the academic publication. \citet{gittelman2003does} explicitly ask ``Does good science lead to valuable knowledge?" in biotechnology. They suggest that ``(…) the evolutionary logics that select valuable scientific publications and valuable patents are different, and because of this, influential publications are not more likely to lead to influential patents than other publications." They employ data on the patent and publication portfolios of 116 biotechnology firms and obtain results that largely confirm their hypothesis.

\citet{suzuki2011structural} argues that patented inventions may be assessed with regard to their monetary or their technical quality. The presence of references to the scientific publications has a strong positive effect on the technological value, but a weak negative effect on the commercial value of the patent. The author also points to considerable heterogeneity across technological fields. \citet{fischer2014testing} use data from Ocean Tomo auctions between 2006 and 2009 to approximate auction prices as a function of observable value correlates. They find only weak and imprecisely estimated effects for the number of NPL references. As they point out, patents traded at Ocean Tomo auctions are not representative and mostly in the IT and IT-related technical fields. \citet{zahringer2017academic} construct a sample of young life science firms and find that higher-quality academic science is associated with patent citations. This relationship is moderated by the respective firm's research activities. \citet{veugelers_scientific_2019} use all Web of Science journal articles published in 2001 and all patents from PATSTAT (version 2013b). They find that only about 10\% of articles become SNPL. Novel publications are more likely to receive future citations by patents, particularly the 1\% highly novel scientific publications. They further find that publications receiving more scientific citations also receive more patent citations.

\citet{sapsalis2006academic} use data on 155 patent families with application dates between 1985 and 1999 at the EPO to model the relationship between citations received by patents and characteristics of the underlying science. They find that NPL self-references (i.e. the inventors are also authors on the referenced scientific publication) to the scientific literature are associated with an increase in forward-citations of a patent. The authors argue that in such cases of highly valuable patents, ``the inventors master (and contribute to) the related science-base (as witnessed by their own publications) and decide to codify their tacit knowledge into technological inventions" \citep[][p.\ 1640]{sapsalis2006academic}. 

In the perspective followed by \citet{fleming2004science}, invention is interpreted as a process of search for new and useful configurations of technological components. Science serves as a map, pointing inventors to particularly useful configurations of components. Alternatively, science allows inventors to avoid search over less productive solutions. However, these effects are not pertinent across all technologies. Recourse to science may offer little help when inventors work with highly independent components, but should generate high returns when the underlying inventive problem is particularly difficult. Using the population of patents granted by the USPTO in May and June of 1990 (n=16,822 after exclusion of 442 patents without any references), they find that only 2,919 of these patents reference scientific publications. In the empirical analysis, the authors show that references to scientific publications increase forward-citations received by patents with an elasticity of about 10\%.

While the results of the studies discussed here are intriguing, they are typically obtained from relatively small samples which are particularly well-suited for the respective studies. An exception is the recent study by \citet{ahmadpoor2017dual}, who analyze the network of US patents citing directly or indirectly SNPL references. They hereby introduce the distance to the science frontier as a metric for science-technology intensity. \citet{Watzinger2018} borrow this metric and provide correlations between the science-technology intensity and the value of patents. \citet{mukherjee2017nearly} emphasize the importance of the age structure of references. The authors study (separately) scientific publications in the WoS database and patents, but they do not link NPL references to WoS entries. Both for publications and for patents they detect a ``hot spot" defined by the age structure (of backward references) that is correlated with an increase in citations received by the publication or patent.

\subsection{Measures of Science Quality}
\label{suppl:sec:measures_science_quality}

\subsubsection*{Scientific citations}
\label{suppl:sec:measures_science_quality_citations}

Our main variable of interest is the scientific quality of publications cited in patents. We use measures of science quality based on the count of forward-citations to publications. This is an established bibliometric indicator of scientific quality. The use of citations is based on the notion that scientists cite publications they consider influential for their own research. Accordingly, it is possible to assume that highly cited publications have a greater impact on follow-on research and represent a meaningful measure of their scientific quality. 

For a given publication, we count the number of citations in a window of three years from publication. This raises the issue that some of these citations may happen later than the filing date of the citing patent. In this case, the number of citations received by a publication may be not independent of the patent itself. In our main specifications, we assume for simplicity that the number of citations to the publication remains indeed independent to the patent citation. In robustness analyses, we verified that the core results remain equivalent when excluding patent citations to publications published in the three years before the filing of the patent.

\subsubsection*{Journal impact factor}
\label{suppl:sec:measures_journal_impact}

An alternative measure of science quality is the impact factor of the journal in which the respective publication is published (JIF). In any given year, the impact factor of a journal is the number of citations, received in that year, of articles published in that journal during the two preceding years, divided by the total number of articles published in that journal during the two preceding years.  We use JIF indicators available by the \href{http://jcr.incites.thomsonreuters.com}{inCite Journal Citations Report}. A disadvantage of this measure is that, due to the lack of completeness of the necessary information, the data are available only after 1997. Moreover, the JIF constitutes a retrospective measure of quality of the journal that ignores the possible high variance of publications quality within one same journal and over time. On the other hand, the JIF has the advantage of being predetermined at the time a publication is published, so that it is not subject to concerns about truncation and mechanical correlation with the measure of patent value.\footnote{We use the JIF as variable of interest in table \ref{reg:pat:sci:imp} to show robustness of our results to alternative measures of science quality.}

\subsubsection*{Patent level aggregation of SNPL references}
\label{suppl:sec:measures_science_quality_aggregation}

In our sample, for patents with SNPL, there are on average \PatFamAvgSNPL \ SNPL references per patent, and a considerable share of \PatFamManySNPLShare\% has references to more than one distinct scientific publication. In our main analyses, we define SNPL science quality as the maximum science quality across publications in SNPL references in a patent. This is based on the notion that the distribution of scientific forward-citations is highly skewed. Consequently, the scientific impact of the most highly cited publication, or the journal with the highest JIF, may be more indicative of SNPL overall science quality than the average across publications. For robustness, we also estimate alternative aggregation operators. This is further discussed in section \ref{suppl:sec:measures_science_quality_aggregation}.\footnote{Table \ref{reg:pat:sci:cit:alt} shows the corresponding results.}

We apply a coherent criterion to aggregate at the patent level the information regarding the presence of self-references: we consider a patent as having a self-reference if the scientific publication with the highest scientific quality among the SNPL references is a self-reference.

\subsection{Measures of Patent Value}
\label{suppl:sec:measures_patent_value}

\subsubsection*{Patent citations}
\label{suppl:sec:measures_patent_value_citations}

Our main dependent variable is patent value. In our main specification, we proxy patent value with the number of forward-citations received by the patent. The number of citations is an established, and perhaps the most widely used, measure of patent value, which is highly correlated with other indicators of technological and economic value of patents \citep{harhoff2003citations,fischer2014testing,moser2018patent}. Patent citations differ substantially from citations in scientific literature. Scientific citations constitute recognition of scientists of the relevance of previous contributions for their own work. In contrast, patent citations, particularly to other patents, perform the legal function of documenting the technological relatedness of a patent to existing prior art with the scope of assessing its novelty and patentability \citep{michel2001patent,roach2013lens}. 

Due to different legal requirements, citations at the EPO and the USPTO differ substantially. EPO patents tend to cite patents that are essential to document the novelty (or lack of novelty) and patentability of the invention; the applicants, in particular, are not required to provide any citation\footnote{Indeed, EPO patents are often filed with no initial references, and, when present, the introduction of references by the applicant is arguably more strategic than in other jurisdictions.}. Applicants at USPTO are expected to report the most extensive list of citations to all possibly relevant patents and examiner complement this list. For this reason we provide analysis where we count EPO and USPTO citations separately.  In our main specifications we use USPTO citations. 

We construct the count of citations to a patent from the USPTO over a period of 5 years from the first filing date.\footnote{The choice of the time window for the count of scientific and patent citations is motivated primarily by pragmatic considerations: we want to ensure a sufficient period so that the number of citations actually reflects the underlying constructs we are interested in, but we want to limit truncation. The difference between the window considered for scientific publications and for patents is also motivated by the fact that patent applications are not instantly published after filing and -- as a result -- typically receive few citations within the first years, whereas scientific publications are often cited immediately after publication.} In robustness analyses, we use the count of citations from the EPO within 5 years from the first filing date. In case of the EPO citation measure, only examiner-supplied citations are considered.

\subsubsection*{Patent scope}
\label{suppl:sec:measures_patent_value_claim}

As alternative proxy for patent value, we adopt a measure of the patent's scope. The value of a patent is considered proportional to the scope of its protection on a particular technology. The narrower the scope of protection, the lower is its value. The text of patent claims tends to be longer for highly specific and narrow patent protection. In other words, longer descriptions of a claimed invention are associated with more specific features that are actually object of the patent protection \citep{kuhn2019scope}. Our measure is defined as the logarithm of the number of words in the first independent claim in patents.\footnote{Tables \ref{reg:pat:sci:cit} and \ref{reg:pat:sci:imp} show the corresponding results. Descriptive statistics are available in the main publication.}

\subsubsection*{Measures of monetary value}
\label{suppl:sec:measures_patent_value_monetary}

Patent citations and patent claim length need to be understood as merely indirect measures of a patent's economic value. Moreover, the number of citations is at times considered to also capture the technological and social value of a patent \citep{trajtenberg1990penny}, which may differ from the private value for the patent owner. Obtaining direct indicators of the monetary private value of patents is a challenging task. Data on this dimension of patent value have limited coverage. To complement the array of indicators of patent value in this direction we adopt two sources of data. First, we use data provided by \citet{kogan2017technological} based on estimated stock market returns to the grant of the patent, as a proxy of the private value of the patent grant. Kogan values are only available for patent families with US patent members where at least one applicant was a publicly listed US company. The data cover exclusively a total of \PatFamXiTotal\, patent families, of which \PatFamXiSNPLTotal\, come with SNPL references. Second, we use survey-based assessments of patent value from the research project PatVal \citep{giuri2007inventors}. This is a subsample of \PatFamPatValTotal\, patent families with at least one EP patent member, of which \PatFamPatValSNPLTotal\, have SNPL references with first filing year mostly in 2003-2005.\footnote{Tables \ref{reg:pat:sci:cit} and \ref{reg:pat:sci:imp} show the corresponding results.} Descriptive statistics are available in the main publication.\footnote{The \citet{kogan2017technological} patent value measures have been in widespread usage since their publication, but in our setting they come with major drawbacks. Much of the private value of the technology will already be incorporated in the stock price, as previous patent publications and grants in other patent systems are informative for investors. The value narrowly captures the additional value of a patent granted in the US patent system. Any information related to the technological capability of the firm that the patent reveals will not be incorporated in that measure. On the other hand, the measures from \citet{giuri2007inventors} are based on a survey, but the exact phrasing measures much more precisely the concept of private patent value: ``Suppose that on the day in which this patent was applied for, the applicant and you had all the information you have today regarding the value of this and the related patents. In case a potential competitor of the applicant was interested in buying the whole set of patents (the patent family including all national patents derived from it), what would have been the minimum price (in Euro) that the applicant should have demanded?''.}

\section{Regression analyses}
\label{suppl:sec:regression}

\subsection{Regression models}
\label{suppl:sec:regression_models}

\subsubsection*{Selection of scientific publications into SNPL references}
\label{suppl:sec:regression_SNPL_probability}

In a first set of analyses, we consider the probability and frequency in which scientific publications appear in SNPL reference, as a function of their scientific quality.

The regressions take the following forms:

\begin{equation}
y_i = \beta_{cit}\ \textit{cit}_i + \sum_{\textit{ft}} \beta_{\textit{ft}} \ \textit{SF}_{\textit{fi}}*T_{\textit{fi}} + \epsilon_i
\end{equation}

\textit{Dependent variable and predictors of interest:} 

\begin{itemize}
    \item $y_i$: The dependent variable is a measure of the probability (or frequency) of a scientific publication appearing among the SNPL references. Respectively, the variable is either a binary or a count variable. Count variables are log-transformed with offset 1. Given the large dataset and the large number of FE groups, nonlinear (count) models are not considered. We employ several variants of these variables.
    \item $\textit{cit}_i$: The main independent variable is a measure of scientific quality. We measure scientific quality at the publication level as the number of citations received over a 3 year period starting from publication (see section \ref{suppl:sec:measures_science_quality}).
\end{itemize}

\textit{FEs:} 

\begin{itemize}
    \item $\textit{SF}_{\textit{fi}}*T_{\textit{fi}}$: These are FEs corresponding to the combination of scientific fields and publication years. These FEs control flexibly for mechanical differences in scientific quality and SNPL frequency across different scientific fields and over time within each scientific field. In total, there are 252 scientific field codes supplied by the Web of Science.
\end{itemize}

\subsubsection*{Science quality and patent value: residualized variables}
\label{suppl:sec:residualization}

Naturally, usage of SNPL references as well as the quality of cited SNPL varies substantially over technological areas as well as over time. In the regression models below, this is taken into account explicitly with FE control variables. In all figures relating patents to scientific quality, we apply residualization which brings the graphical display in line with the regression outputs.

To do so, we regress both the SNPL science quality variables as well as the patent value variables on the full set of technology area $\times$ first filing year FEs. The formal model reads $y_i = \sum_{\textit{ft}} \beta_{\textit{ft}}F_{\textit{fi}}*T_{\textit{ti}} + \epsilon_i$. This is done in the full sample of patents both with and without SNPL references. Afterwards, we calculate the residual variation as $\hat{\epsilon}_i \equiv y_i - \hat{y}_i = y_i - \sum_{\textit{ft}} \hat{\beta}_{\textit{ft}} F_{\textit{fi}}*T_{\textit{ti}}$, where $\hat{\epsilon}$, $\hat{y}$ and $\hat{\beta}$ are estimated values. In fact, $\hat{\epsilon}_i = y_i - \bar{y}_{\textit{ft}}$, where $\bar{y}_{\textit{ft}}$ is the mean within technology area $\times$ first filing year group. Therefore, $E[\hat{\epsilon}_i] = 0$, both overall and within each $\textit{ft}$ group.

The values plotted in the graphs are $\hat{\epsilon}_i + \bar{y}$, where $\bar{y}$ is the full-sample mean of $y$. This returns the absolute levels back to what is contextually expected and interpretable.

In plain terms, this strategy removes level effects within technology area $\times$ first filing year groups by subtracting the mean $y$ within groups. The overall level is retained by adding the overall $y$ mean. The $y$ variable is transformed. Before, it is a deviation from the \textit{within-group} mean. Afterwards, it is a deviation from the \textit{overall} mean.

\subsection*{Science quality and patent value: regression models}
\label{suppl:sec:regression_science_quality}

In the empirical analysis, we study the relationship between the presence and the quality of scientific publications referenced in patents and the value of patents. 

The regressions take the following form:
\begin{equation}
\begin{split}
y_i &= \beta_{\textit{hasSNPL}} \ \textit{hasSNPL}_i + \beta_{\textit{snplQ}} \ \textit{snplQ}_i \\[8pt]
&+ \sum_{\textit{ft}} \beta_{\textit{ft}}TF_{\textit{fi}}*T_{\textit{ti}} + \sum_{a} \beta_{a}A_{\textit{ai}} + \sum_{n} \beta_{n}N_{\textit{ni}} + \sum_{r} \beta_{r}R_{\textit{ri}} + \sum_{p} \beta_{p}P_{\textit{pi}} + \epsilon_i
\end{split}
\end{equation}

\textit{Dependent variable and predictors of interest:} 
\begin{itemize}
\item $y_i$: The dependent variable is a measure of patent value. In the main specifications and figures, we use the count of citations from the USPTO within the first 5 years after filing. In alternative specifications, we use: the count of citations from the EPO; indicators of monetary value; patent scope as measured by the length of the first independent claim (see section \ref{suppl:sec:measures_patent_value}). All dependent variables are in log-terms with offset 1. Given the large dataset and the large number of FE groups, nonlinear (count) models could not be considered.
\item $\textit{hasSNPL}_i$: A dummy equal to 1 if a patent has at least one reference to a scientific publication
\item $\textit{snplQ}_i$: A measure of SNPL science quality. We measure scientific quality at the scientific publication level as the number of citations received over a period of 3 years from publication. We define SNPL science quality as the maximum scientific quality across SNPL references in a patent when more than one is present.\footnote{We test the robustness of the results to alternative aggregation criteria (see table \ref{reg:pat:sci:cit:alt}).}
\end{itemize}

\textit{FEs:} 
\begin{itemize}
\item $TF_{\textit{fi}}*T_{\textit{ti}}$: These are FEs corresponding to the combination of technological classes and first filing year. These FEs control flexibly for mechanical differences in patent value across different technological fields and over time within each technological field.
\item $A_{\textit{ai}}$: These are FEs for the applicant of the patent. 
\item $N_{\textit{ni}}$: These are FEs for the distinct number of inventors listed on the patent.
\item $R_{\textit{ri}}$: These are FEs for the number of patent references. We use individual FEs for each number of references up to the 95th percentile and assign one dummy for all patents with a higher number of references.\footnote{In regressions involving PatVal (EUR) values, the number of available observations is substantially lower. Here, we include only the log-transformed count of backward patent references when estimating the extended specification.}
\item $P_{\textit{pi}}$: These are FEs for the number of patent references to scientific publications. We use an individual FE for each number of references up to the number corresponding to the 95th percentile and aggregate in one FEs patents with a higher number of references. Note that $\textit{hasSNPL}_i$ is collinear and therefore dropped when this set of FEs is used.
\end{itemize}

\subsection{Regression results}
\label{suppl:sec:regression_results}

\subsubsection*{Selection of scientific publications in SNPL references} 

\begin{figure}[htb]
\centering
\caption{Selection into SNPL by science quality}
\label{supp:fig:paper:selection}
\begin{subfigure}{.5\textwidth}
    \centering
    \caption{Probability of patent citation (ext.\ margin) \label{supp:fig:paper:percentiles:p}}
        \includegraphics[width=1\linewidth]{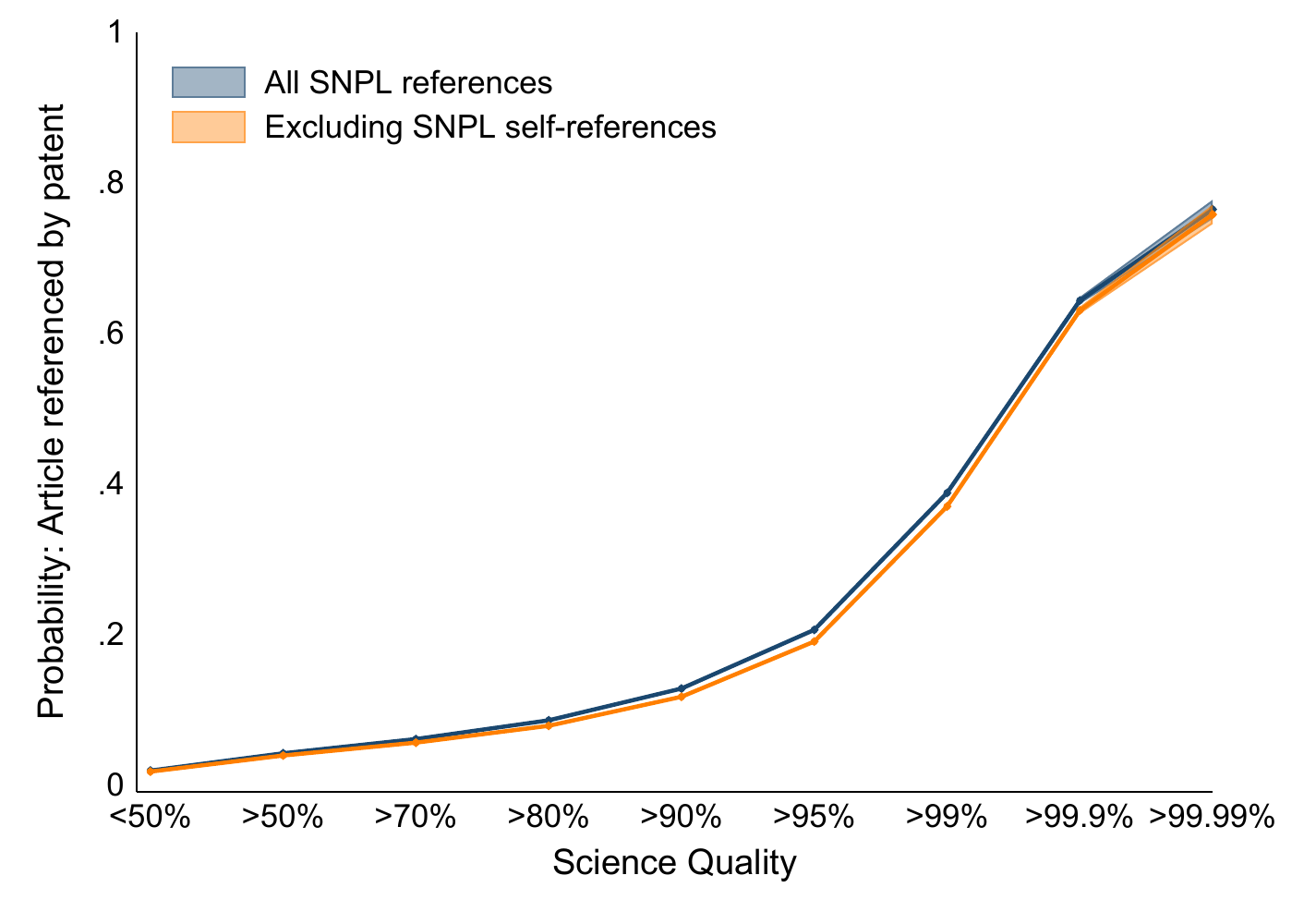}
\end{subfigure}%
\begin{subfigure}{.5\textwidth}
    \centering
    \caption{Number of patent citations (int.\ margin) \label{supp:fig:paper:percentiles:n}}
        \includegraphics[width=1\linewidth]{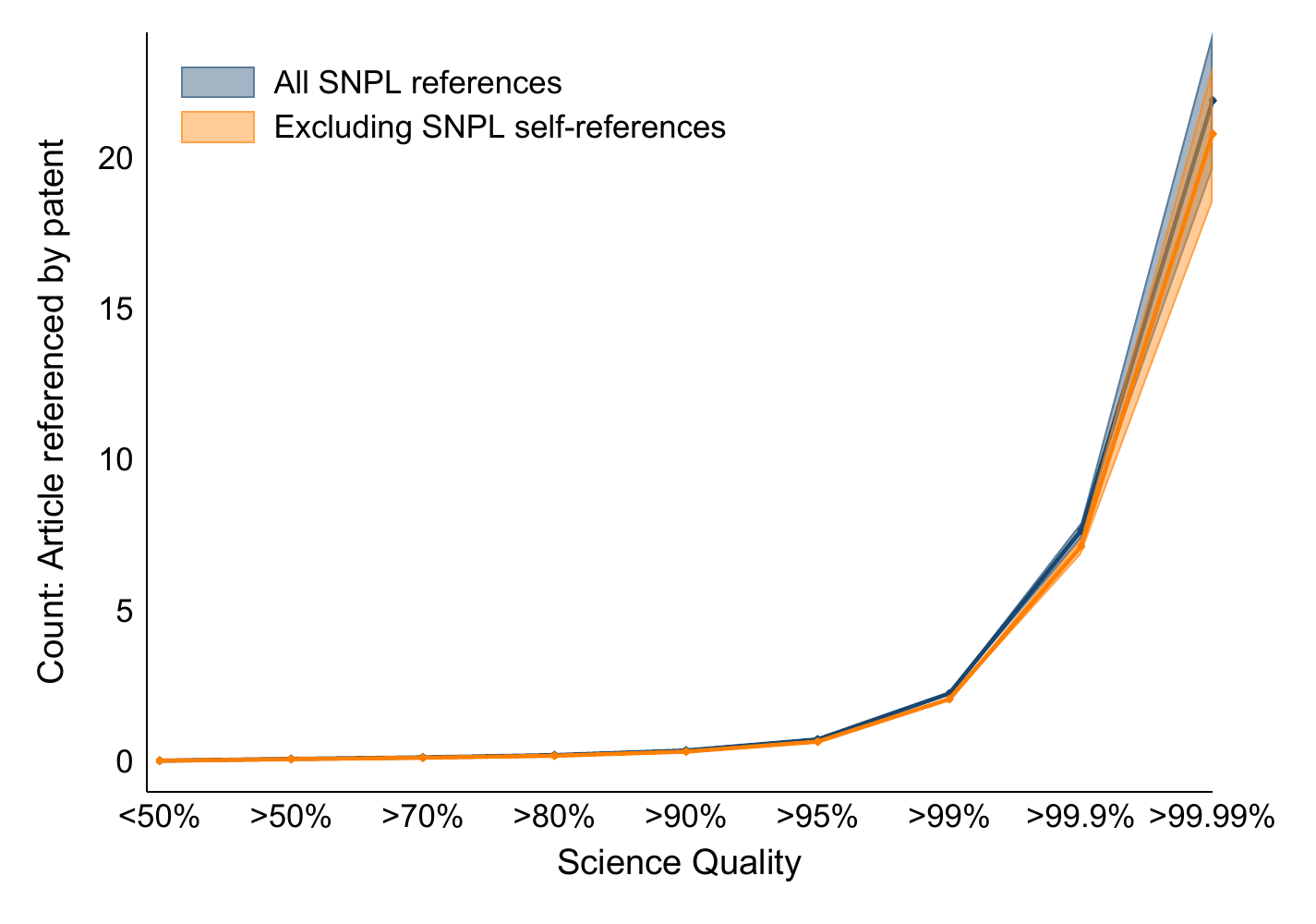}
\end{subfigure}
    \legend{\footnotesize{\textbf{Notes:} Probability of being cited as SNPL (left) and the number of SNPL citations (right) by science quality of a scientific publication. Science quality is the 3-year citation count of the scientific publication. Shaded areas show 95\% confidence intervals around the respective means.
}}
\end{figure}

We present first regression results for the probability that a scientific publication appears in SNPL references as a function of its scientific quality. In the first main specification, table \ref{reg:select:elasticities}, column 1 and 2, we consider all SNPL references. Second, in column 3 and 4, we consider exclusively SNPL references within five years from the year of publications. Third, in column 5 and 6, we consider references within five years and exclusively if they are the SNPL references with the highest scientific quality. In a fourth variant, column 7 and 8, we consider only SNPL references that are cited for the first time by an applicant, so that each patent applicant-scientific publication pair is counted at most once (\textit{one per applicant}). Finally, in table \ref{reg:select:elasticities:restrict}, we provide regression results excluding academic patents as well as self-references of various types. Figures \ref{supp:fig:paper:percentiles:p} and \ref{supp:fig:paper:percentiles:n} also show graphically that the exclusion of SNPL self-references is irrelevant to the results. Overall, we consistently find a positive and significant effect of science quality on the selection of scientific articles into SNPL references.

\subsubsection*{Main regression results: SNPL science quality and patent value} 

Table \ref{reg:pat:sci:cit} presents regression results for our core findings. It shows elasticity estimates for the main measure of SNPL science quality and each one of the alternative measures of patent value as dependent variable. We include sets of more demanding controls incrementally. All models include the variable $\textit{hasSNPL}_i$ as a control for the level effect of having at least one SNPL reference. In column 1 to 6, we present results for our base-line specification where we control exclusively for technology field and year pair FEs. In column from 7 to 12, we include all patent level controls as detailed in the above section \ref{suppl:sec:regression_models}. In column from 13 to 18, we add applicant FEs. Figure \ref{supp:fig:paper:has_npl} further highlights the striking differences in the overall distribution of patent citations for patents with and without SNPL references. 

\subsubsection*{Alternative measures of SNPL science quality} 

\begin{figure}[htb]
    \begin{subfigure}{.5\textwidth}
    \centering
        \includegraphics[width=.8\linewidth]{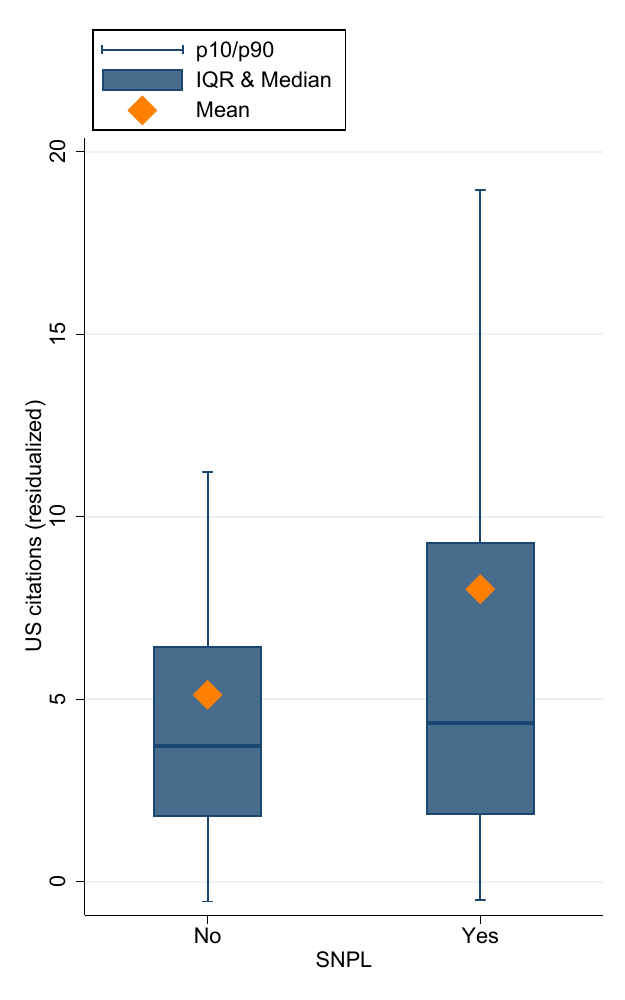}
        \caption{Patent value by science reference \label{supp:fig:paper:has_npl}}
    \end{subfigure}
    \begin{subfigure}{.5\textwidth}
        \centering
        \includegraphics[width=\linewidth]{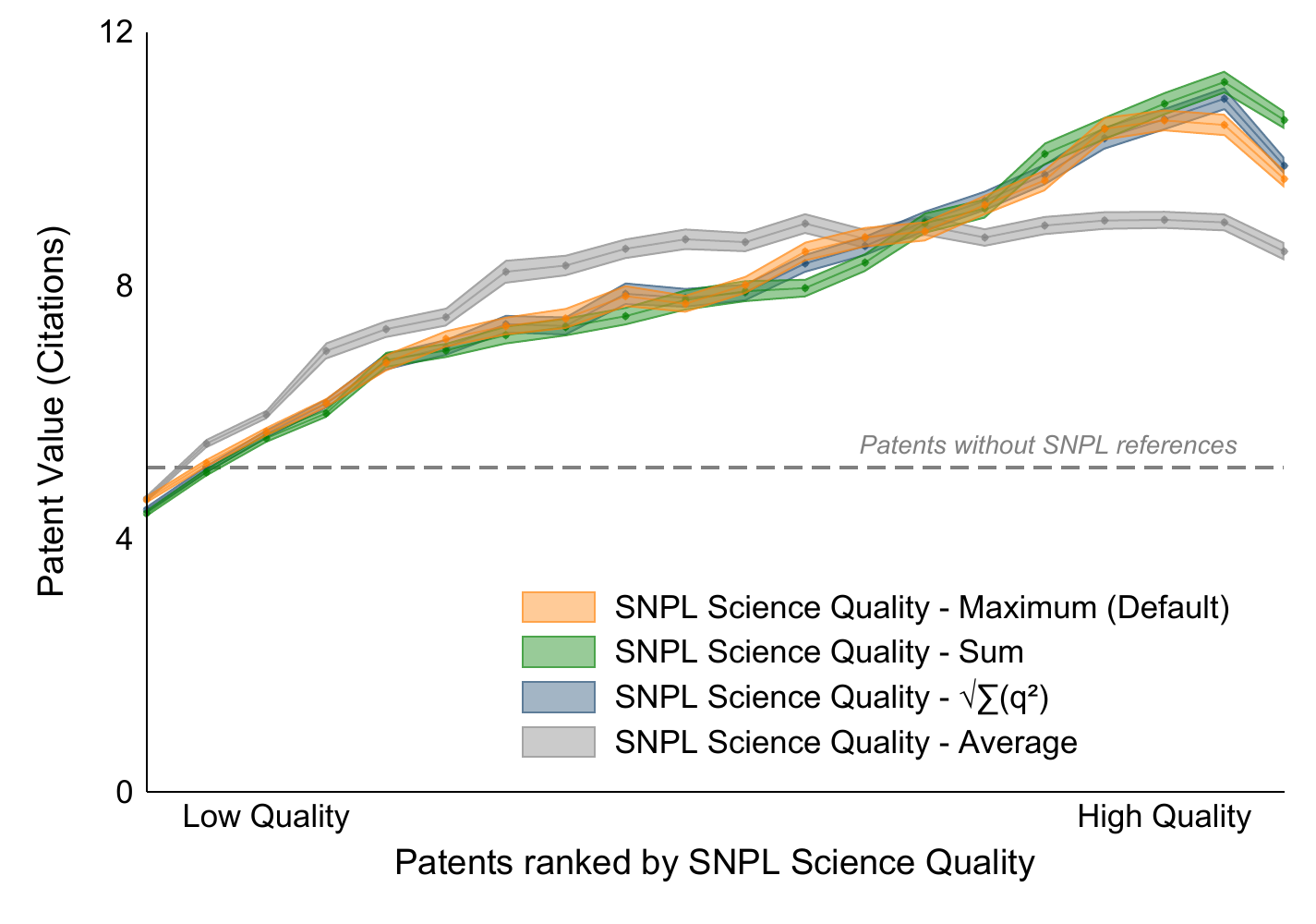}
        \caption{Patent value by SNPL science quality \label{supp:fig:paper:percentiles}}
    \end{subfigure}
    \legend{\footnotesize{\textbf{Notes:} Left: Distribution of patent citations for patents with and without SNPL references. Residualized 5-year patent forward-citations by US patents towards patent families are used, see section \ref{suppl:sec:residualization}.

    Right: Average patent values by science quality, considering alternative science quality operationalizations. SNPL science quality is the quality of publications referenced by a patent. When there are multiple patent-paper references, we by default use the highest-quality reference (orange). In comparison, the average quality also delivers a positive correlation (gray), but it is more diluted. Other aggregation methods which also focus on the top of the distribution are virtually identical to the maximum. These are the sum (green) and the square root of the sum of squares (blue). Science quality is the 3 year citation count of a scientific publications. Patent value is measured as the 5 year count of patent forward citations by US patents. Patent value and science quality are residualized using technology field $\times$ first filing year FEs. The dashed line indicates the average patent value of patents without SNPL references. Shaded areas show 95\% confidence intervals around the respective means. N = \PatFamTotal{} patents (\PatFamSNPLTotal{} with SNPL references).}}
\end{figure}

As a first variant to these specifications, we test the robustness of the results to alternative measures of SNPL science quality. In table \ref{reg:pat:sci:imp} we use a measure based on the journal impact factor instead of citations. The number of observations is lower because the journal impact factors are only available to us from 1998 onward. Overall, we find very similar results.\footnote{The only exception are the results for the USD values, where the results are unstable in the first two specifications (column 5 and 11) but remain positive and significant in our last and most complete specification.} In table \ref{reg:pat:sci:cit:alt}, we use alternative measures of SNPL science quality derived from different criteria of aggregation at the patent level of the scientific quality of multiple scientific publications, when more than one appear in the NPL-references of a patent. When $c_i$ is the citation count of SNPL reference $i$, in our main models we consider the maximum. Alternatively, we also consider the sum ($\sum_i c_i$), average ($\frac{1}{n} \sum_i c_i$) and square root of the sum of squares ($\sqrt {\sum_i c_i^2}$). We find similar results irrespective of the aggregation criterion used. Figure \ref{supp:fig:paper:percentiles} graphically shows the results.

\subsubsection*{Self-references} 

\begin{figure}[htb]
\centering
\caption{Patent value by science quality, with and without SNPL self-references}
\captionsetup[subfigure]{justification=centering}
\begin{subfigure}{.50\textwidth}
    \centering
    \caption{Share of SNPL (self-)references by \\ science quality \label{supp:fig:paper:self:desc}}
        \includegraphics[width=1\linewidth]{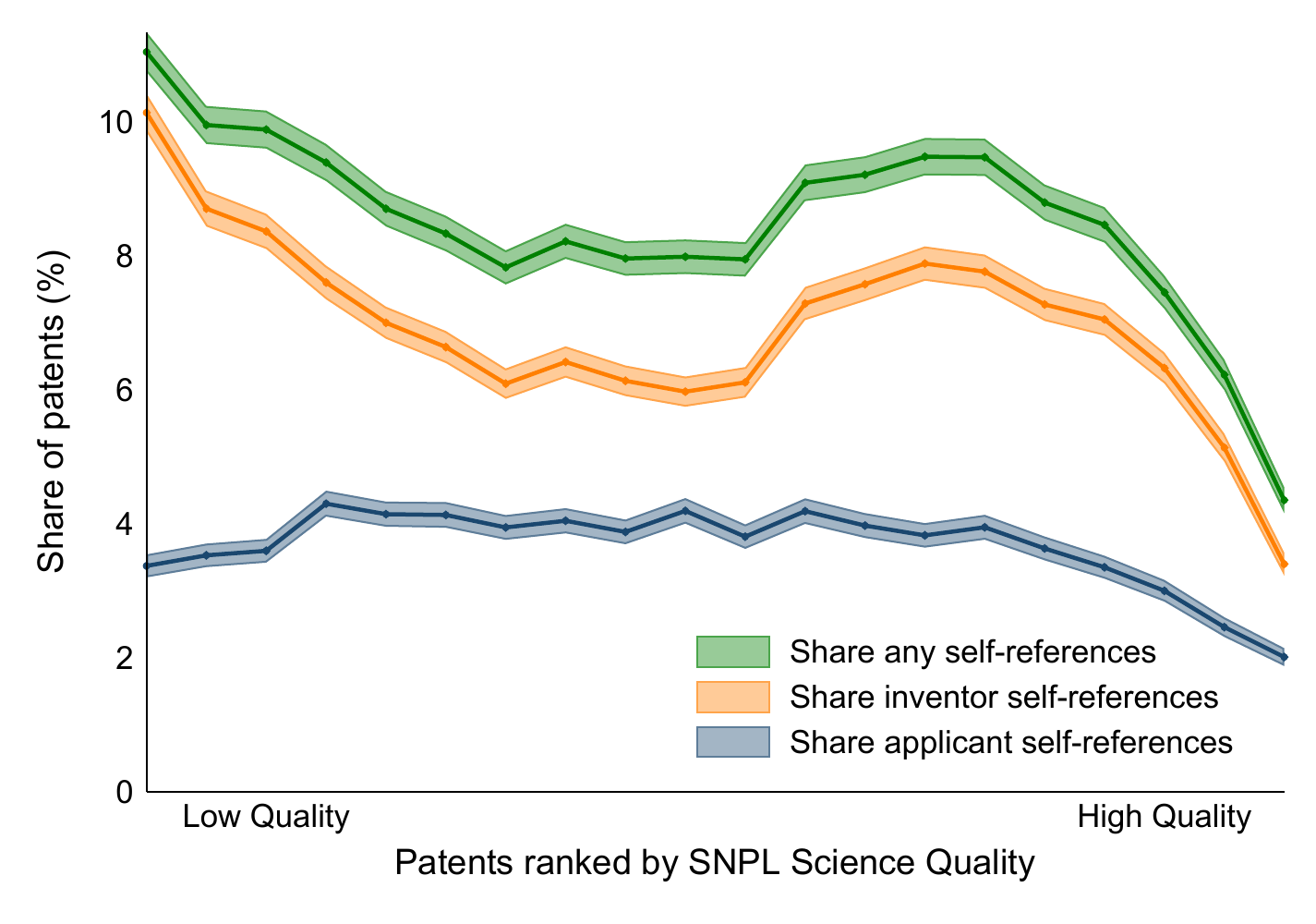}
\end{subfigure}%
\begin{subfigure}{.50\textwidth}
    \centering
    \caption{Patent value by science quality of \\ SNPL (self-)references \label{supp:fig:paper:self}}
        \includegraphics[width=1\linewidth]{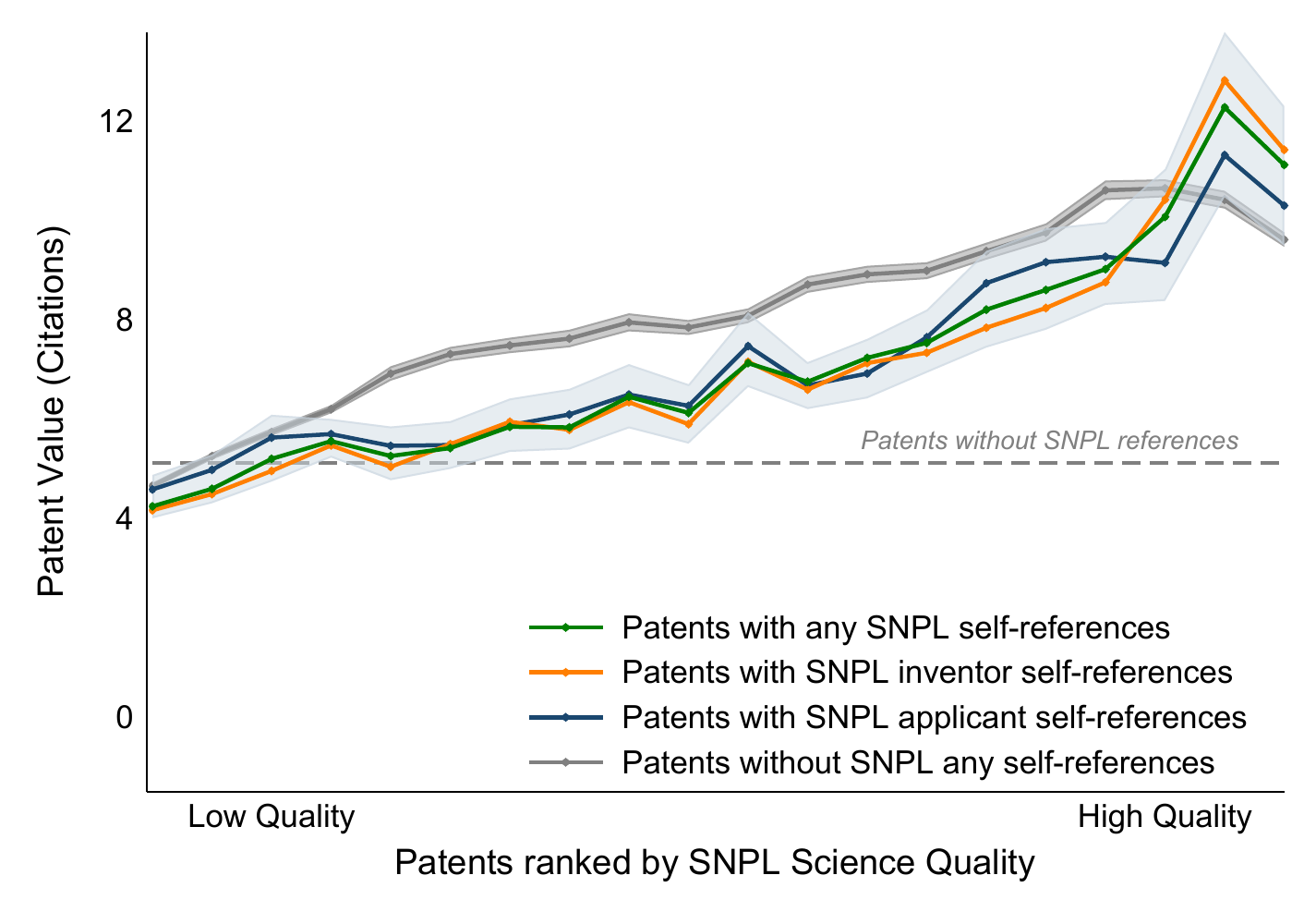}
\end{subfigure}
        \legend{\footnotesize{\textbf{Notes:} Left: Share of SNPL self-references by SNPL science quality. Right: Average patent value by SNPL science quality and categories of SNPL self-references. The lines show values for any self-reference (green), inventor self-references (orange) and applicant self-references (blue) or patents without SNPL self-references (gray). SNPL science quality is the maximum 3-year citation count across scientific publications appearing as SNPL references in a patent. Patent value is measured as the 5-year count of patent forward-citations by US patents. Patent value and science quality measures are residualized using technology field-first filing year pair FEs. The gray shaded area shows 95\% confidence intervals around the respective means. For visual purposes, the confidence intervals around the means concerning self-references show the maximum extent of the 95\% confidence intervals of any of the three underlying measures.
}}
\end{figure}

Figure \ref{supp:fig:paper:self:desc} shows the frequency of occurrence of self-references: between 5 and 10\% of all patent families include a self-reference. Most self-references are inventor self-references (5-10\%), whereas applicant self-references are less frequent with 2-4\%. The frequency of self-references tends to decrease with the SNPL science quality (although non-monotonically); this tendency is most pronounced at the top. 

In the paper, we consider the possibility that self-references drive the results. On the one hand, from a theoretical standpoint, it is interesting to consider whether high-quality science leads to high-value technologies within or outside the boundaries of the organizations in which it is developed. On the other hand, we want to ensure that the results are not driven by highly productive organizations and individuals that perform scientific and technological activities at the same time. 

Figure \ref{supp:fig:paper:self} replicates the results reported in the paper, separating different categories of self-references. The different groups of self-references behave very similarly. Table \ref{reg:pat:sci:cit:self} provides regression estimates separately for a sample consisting only of patents with self-references and excluding all patents with self-references. While the magnitude is larger for the sample with self-references, the estimated elasticities are positive and significant in all specifications. We can conclude that self-references do not drive the overall effects. 

\subsubsection*{Technology fields, year of patent filing, and applicant countries} 

We analyze the heterogeneity of the estimates, first, across technological fields of patents. In table \ref{reg:pat:sci:cit:area}, we run separate regressions by technology main area. In line with previous literature \citep[e.g.,][]{harhoff2003citations}, we find that effects are particularly strong in Chemistry. However, SNPL science quality also matters for Electrical Engineering, Instruments as well as Mechanical Engineering.

\begin{figure}[htb]
    \centering
    \caption{Patent value-science quality relationship over time\label{supp:fig:cite:UScit:time:max}}
        \begin{subfigure}{.45\linewidth}
        \caption{Baseline regressions}      
        \includegraphics[width=\linewidth]{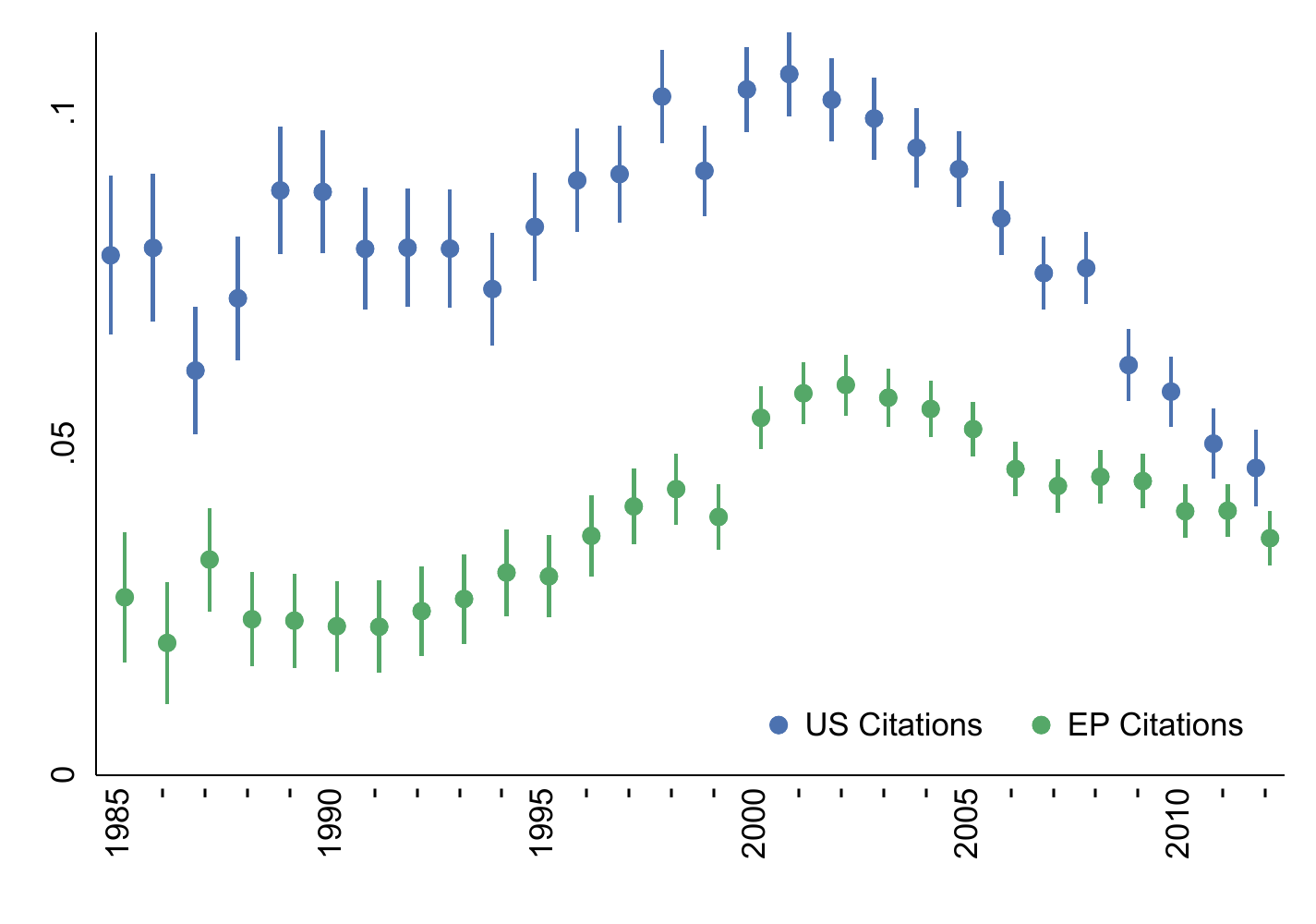}
        \end{subfigure}%
        \begin{subfigure}{.45\linewidth}
        \caption{All patent-level controls}        
        \includegraphics[width=\linewidth]{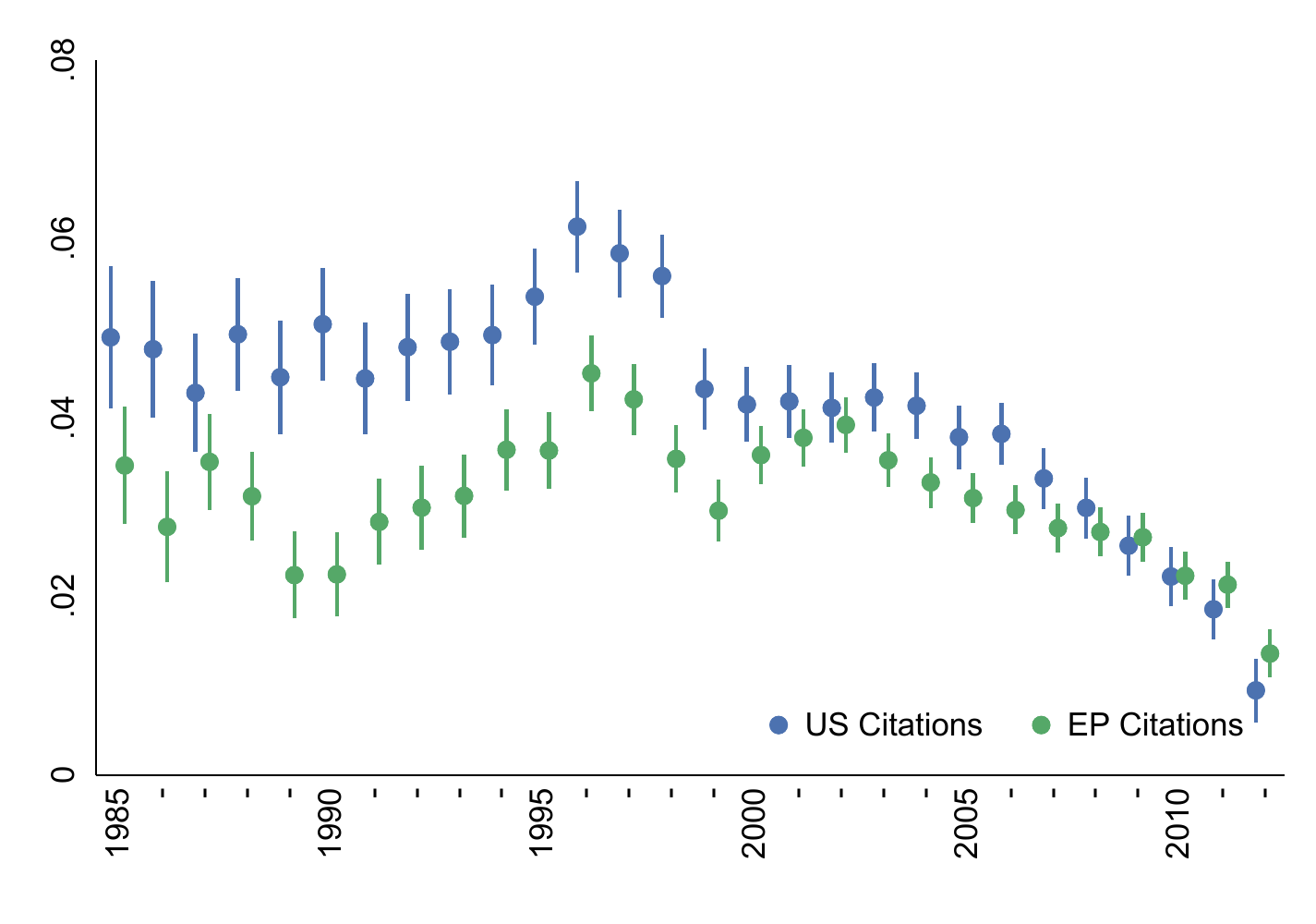}
        \end{subfigure}
    \legend{\footnotesize{\textbf{Notes:} The figure plots the interactions coefficients between each first filing year of patents and SNPL science quality, in a regression with the 5-year patent forward-citations by US patents and EP patents as dependent variables. Science quality is the maximum 3-year citation count of SNPLs of a patent family. Left: Models include technology field and first filing year pair FEs. Right: Models additionally include FEs for SNPL reference counts, patent reference counts and number of inventors. Range indicators show 95\% confidence intervals around the respective regression coefficients.
      
      }}
\end{figure}

Second, we explore the heterogeneity over time based on the first filing year of patents. We decompose the elasticities calculated in table \ref{reg:pat:sci:cit} (column 1/2, 7/8) over time. Figure \ref{supp:fig:cite:UScit:time:max} depicts the corresponding point estimates. We find that after marginally increasing between 1985 and 2000, the extent of the relationship decreased substantially. After 2000, there is a substantial decline which is especially pronounced for the US system. The understanding of the reasons for this decline requires further research.

We consider the possibility that the effect is driven by intense science usage of particular countries. To do so, we split the sample by the first applicant country and consider China, Europe (EU-28), Japan, South Korea and the United States separately. From table \ref{reg:pat:sci:cit:ctry}, we find that science quality is important in all countries, but particularly so in Europe, the US and Japan. Overall, the results are consistent across different geographic areas.

\subsubsection*{Interdisciplinarity of SNPL references} 

We explore the role of interdisciplinarity of science. Previous studies demonstrate the existence of a close connection between novelty and scientific impact and the ability of scientists to successfully recombine knowledge from distinct domains \citep{mukherjee2017nearly,wang2017bias,veugelers_scientific_2019}. In the context of our analysis we are interested in exploring whether this dimension explains the correlation between SNPL science quality and patent value. We proxy the interdisciplinarity of science with the interdisciplinary journals as captured by the classification of journals in scientific fields in WoS.\footnote{Some field codes refer directly to multidisciplinary research. These field codes are ah, vj, wu, bq, po, ev, ui, dy, le, ro, if and pm. We tested our results by including journals associated with these codes in the sample of interdisciplinary journals or excluding them. This affects the level estimates of patent value for different values of interdisciplinarity but leaves untouched the correlation with SNPL science quality.}

\begin{figure}[htb]
    \caption{Patent value by science quality and interdisciplinarity \label{supp:fig:interdisciplinarity}}
    \centering
    \begin{subfigure}{.5\textwidth}
    \centering
    \caption{Share of interdisciplinarity \label{supp:fig:paper:interdisciplinarity:share}}
        \includegraphics[width=\linewidth]{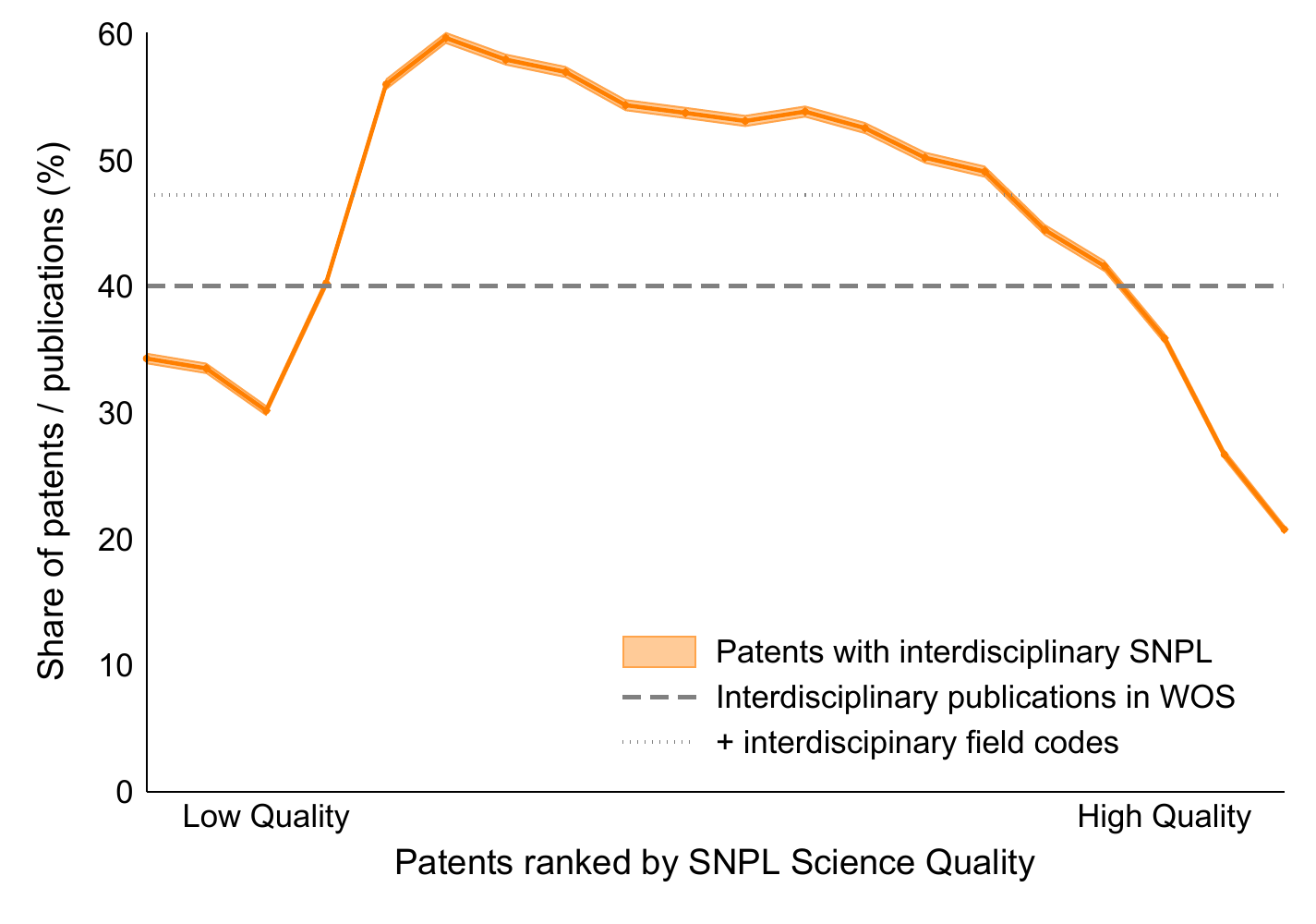}
    \end{subfigure}%
    \begin{subfigure}{.5\textwidth}
        \centering
        \caption{Patent value and interdisciplinarity \label{supp:fig:paper:interdisciplinarity}}
        \includegraphics[width=\linewidth]{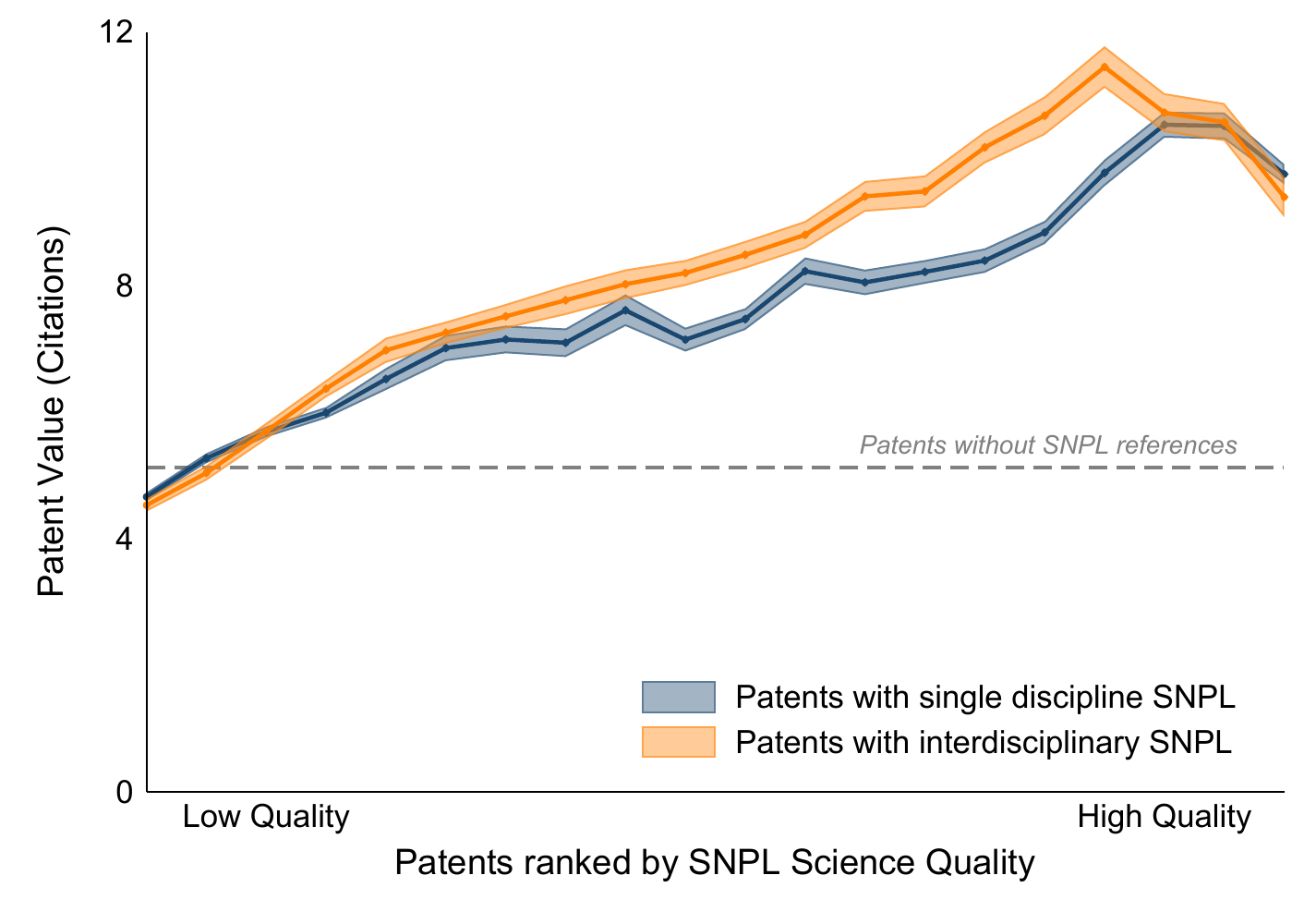}
    \end{subfigure}
    \legend{\footnotesize{\textbf{Notes:} Left: Share of patents with interdisciplinary SNPL by SNPL science quality. Right: Patent value by SNPL science quality and by interdisciplinarity of SNPL references. Scientific articles are considered interdisciplinary if the journal where they are published is associated with at least two WoS field codes. SNPL science quality is the maximum 3-year citation count across scientific publications appearing as SNPL references in a patent. Patent value is measured as the 5-year count of patent forward-citations by US patents. Patent value and science quality measures are residualized using technology field-first filing year pair FEs. Shaded areas show 95\% confidence intervals around the respective means.
}}
\end{figure}

In figure \ref{supp:fig:paper:interdisciplinarity:share} we plot the share of patents with SNPL references to interdisciplinary scientific publications and, in figure \ref{supp:fig:paper:interdisciplinarity}, the patent value by SNPL science quality of patents with and without SNPL interdisciplinary references. The share of patents with interdisciplinary SNPL references is highest for intermediary values of SNPL science quality. Indeed, we find that interdisciplinarity is associated overall with higher patent value, with the exception of patents at the top of the SNPL science quality distribution. The correlation with SNPL science quality remains in any case highly positive for both categories. Table \ref{reg:pat:sci:cit:inter} presents the underlying regression results.

\subsubsection*{Distance of SNPL references} 

Finally, we present regression results for the distance and time-distance of SNPL references. The related results are presented graphically in the paper. Table \ref{reg:pat:sci:cit:frontier} shows results relative to the interaction between the distance of SNPL references and table \ref{reg:pat:sci:cit:lag} reports the corresponding regression results for the time-distance of SNPL references. Here, we split the time-distance into tertiles. In accordance to what is discussed in the paper, we find that patent families at a short distance, by either dimension, are of particularly high value and tend to show higher elasticities with SNPL science quality as well. The elasticities remain in any case strongly positive and significant also at a relatively high distance. 
\clearpage

\section{Supplementary graphs and tables}
\label{suppl:sec:suppl_graphs_tables}

\clearpage

\begin{landscape}
\section{Regression results}
\begin{table}[h]
\setlength{\tabcolsep}{8pt}
        \caption{SNPL and science quality elasticities (intensive and extensive margin, by SNPL definitions) \label{reg:select:elasticities}}
        \centering
        \small
 \renewcommand{\arraystretch}{1.2}  
        \begin{tabular}{@{}l*{8}{S}@{}} \toprule
SNPL definition                    &\multicolumn{2}{c}{{All}}&\multicolumn{2}{c}{{Within 5y}}&\multicolumn{2}{c}{{Within 5y max quality}}&\multicolumn{2}{c}{{One per applicant}}\\
                    &\multicolumn{1}{c}{(1)}&\multicolumn{1}{c}{(2)}&\multicolumn{1}{c}{(3)}&\multicolumn{1}{c}{(4)}&\multicolumn{1}{c}{(5)}&\multicolumn{1}{c}{(6)}&\multicolumn{1}{c}{(7)}&\multicolumn{1}{c}{(8)}\\
DV: SNPL            &     {(1/0)}&     {Count}&     {(1/0)}&     {Count}&     {(1/0)}&     {Count}&     {(1/0)}&     {Count}\\
\midrule
3y Cit              &       0.053&       0.068&       0.042&       0.049&       0.013&       0.013&       0.041&       0.043\\
                    &   (1550.99)&   (1575.22)&   (1405.31)&   (1435.04)&    (785.91)&    (820.06)&   (1398.84)&   (1470.63)\\
\midrule
Field $\times$ Year FE&     {{Yes}}&     {{Yes}}&     {{Yes}}&     {{Yes}}&     {{Yes}}&     {{Yes}}&     {{Yes}}&     {{Yes}}\\
Adj.\ R-Square      &     {0.114}&     {0.111}&     {0.091}&     {0.090}&     {0.033}&     {0.033}&     {0.091}&     {0.094}\\
Observations        &  {42962463}&  {42962463}&  {42962463}&  {42962463}&  {42962463}&  {42962463}&  {42962463}&  {42962463}\\
\bottomrule
\end{tabular}

      \legend{\footnotesize{\textbf{Notes:} Values in ``0/1''-columns are semi-elasticities, values in ``Count''-columns are elasticities. Includes WoS subject code times publication year FEs. The level of observation is at the WoS item. Science quality (\textit{3y cit}) is measured by 3-year forward-citations by other WoS items. Robust standard errors. T-statistics in parentheses.}}
\vspace{0.5cm}
\setlength{\tabcolsep}{8pt}
        \caption{SNPL and science quality elasticities (probability and frequency, with SNPL restrictions)  \label{reg:select:elasticities:restrict}}
        \centering
        \small
 \renewcommand{\arraystretch}{1.2}  
        \begin{tabular}{@{}l*{8}{S}@{}} \toprule
SNPL restriction                    &\multicolumn{2}{c}{{No academic patents}}&\multicolumn{2}{c}{{No applicant self-ref.}}&\multicolumn{2}{c}{{No inventor self-ref.}}&\multicolumn{2}{c}{{No any self-ref.}}\\
                    &\multicolumn{1}{c}{(1)}&\multicolumn{1}{c}{(2)}&\multicolumn{1}{c}{(3)}&\multicolumn{1}{c}{(4)}&\multicolumn{1}{c}{(5)}&\multicolumn{1}{c}{(6)}&\multicolumn{1}{c}{(7)}&\multicolumn{1}{c}{(8)}\\
DV: SNPL            &     {(1/0)}&     {Count}&     {(1/0)}&     {Count}&     {(1/0)}&     {Count}&     {(1/0)}&     {Count}\\
\midrule
3y Cit              &       0.036&       0.046&       0.052&       0.067&       0.050&       0.063&       0.050&       0.063\\
                    &   (1208.51)&   (1214.60)&   (1536.94)&   (1558.94)&   (1502.40)&   (1518.22)&   (1498.92)&   (1514.58)\\
\midrule
Field $\times$ Year FE&     {{Yes}}&     {{Yes}}&     {{Yes}}&     {{Yes}}&     {{Yes}}&     {{Yes}}&     {{Yes}}&     {{Yes}}\\
Adj.\ R-Square      &     {0.082}&     {0.078}&     {0.112}&     {0.109}&     {0.108}&     {0.104}&     {0.107}&     {0.103}\\
Observations        &  {42962463}&  {42962463}&  {42962463}&  {42962463}&  {42962463}&  {42962463}&  {42962463}&  {42962463}\\
\bottomrule
\end{tabular}

      \legend{\footnotesize{\textbf{Notes:} Includes WoS subject code times publication year FEs.  The level of observation is at the WoS item. Science quality (\textit{3y cit} is measured by 3-year forward-citations by other WoS items. The baseline category consists of observations with no 3-year forward-citations, approximately 50\% of the dataset. Robust standard errors. T-statistics in parentheses.}}
\end{table}
\end{landscape}

\begin{table}[ht]
\setlength{\tabcolsep}{4pt}
        \caption{Patent value and science quality \label{reg:pat:sci:cit}}
        \centering
        \small
 \renewcommand{\arraystretch}{1.1}  
        \begin{tabular}{@{}l*{6}{S}@{}} \toprule
                    &\multicolumn{1}{c}{(1)}&\multicolumn{1}{c}{(2)}&\multicolumn{1}{c}{(3)}&\multicolumn{1}{c}{(4)}&\multicolumn{1}{c}{(5)}&\multicolumn{1}{c}{(6)}\\
DV (log):           & {5y Cit US}& {5y Cit EP}&{US Claim Length}&{EP Claim Length}&{USD Values}&{EUR Values}\\
\midrule
3y Cit SNPL ref (max)&       0.082&       0.042&      -0.015&      -0.059&       0.022&       0.107\\
                    &    (123.51)&     (86.00)&    (-27.42)&    (-39.51)&      (6.83)&      (3.47)\\
\midrule
Patent-level controls&    {{Base}}&    {{Base}}&    {{Base}}&    {{Base}}&    {{Base}}&    {{Base}}\\
Patent applicant FE &      {{No}}&      {{No}}&      {{No}}&      {{No}}&      {{No}}&      {{No}}\\
Adj.\ R-Square      &     {0.156}&     {0.067}&     {0.157}&     {0.323}&     {0.113}&     {0.045}\\
Observations        &   {4319309}&   {4319309}&   {2464480}&   {1241037}&    {899272}&     {10839}\\
\bottomrule
\end{tabular}
\begin{tabular}{@{}l*{8}{S}} & & \\ \toprule &\multicolumn{1}{c}{{(7)}} &\multicolumn{1}{c}{{(8)}} &\multicolumn{1}{c}{{(9)}} &\multicolumn{1}{c}{{(10)}} &\multicolumn{1}{c}{{(11)}}&\multicolumn{1}{c}{{(12)}}\\
DV (log):           & {5y Cit US}& {5y Cit EP}&{US Claim Length}&{EP Claim Length}&{USD Values}&{EUR Values}\\
\midrule
3y Cit SNPL ref (max)&       0.037&       0.030&      -0.012&      -0.038&      -0.046&       0.082\\
                    &     (45.46)&     (47.43)&    (-17.10)&    (-18.79)&    (-11.27)&      (2.69)\\
\midrule
Patent-level controls&     {{All}}&     {{All}}&     {{All}}&     {{All}}&     {{All}}&     {{All}}\\
Patent applicant FE &      {{No}}&      {{No}}&      {{No}}&      {{No}}&      {{No}}&      {{No}}\\
Adj.\ R-Square      &     {0.262}&     {0.100}&     {0.160}&     {0.324}&     {0.125}&     {0.065}\\
Observations        &   {4319309}&   {4319309}&   {2464480}&   {1241037}&    {899272}&     {10839}\\
\bottomrule
\end{tabular}
\begin{tabular}{@{}l*{8}{S}} & & \\ \toprule &\multicolumn{1}{c}{{(13)}} &\multicolumn{1}{c}{{(14)}} &\multicolumn{1}{c}{{(15)}} &\multicolumn{1}{c}{{(16)}} &\multicolumn{1}{c}{{(17)}}&\multicolumn{1}{c}{{(18)}}\\
DV (log):           & {5y Cit US}& {5y Cit EP}&{US Claim Length}&{EP Claim Length}&{USD Values}&{EUR Values}\\
\midrule
3y Cit SNPL ref (max)&       0.027&       0.023&      -0.010&      -0.027&       0.003&       0.084\\
                    &     (29.68)&     (31.57)&    (-12.13)&    (-13.00)&      (1.94)&      (1.52)\\
\midrule
Patent-level controls&     {{All}}&     {{All}}&     {{All}}&     {{All}}&     {{All}}&     {{All}}\\
Patent applicant FE &     {{Yes}}&     {{Yes}}&     {{Yes}}&     {{Yes}}&     {{Yes}}&     {{Yes}}\\
Adj.\ R-Square      &     {0.362}&     {0.169}&     {0.253}&     {0.389}&     {0.887}&     {0.113}\\
Observations        &   {3763831}&   {3763831}&   {2099196}&   {1122794}&    {857176}&      {5697}\\
\bottomrule
\end{tabular}

      \legend{\footnotesize{\textbf{Notes:} All reported values are elasticities.
      \textit{3y Cit SNPL ref (max)} is a measure of SNPL science quality corresponding to the maximum 3-year citation count across scientific publications appearing as SNPL references in a patent. Patent-level controls ``Base'' include technology fields and first filing year pair FEs. Patent-level controls ``All'' further include FEs for SNPL reference counts, patent reference counts and number of inventors. Patent applicant FEs are based on the first applicant on the grant publication. Robust standard errors. T-statistics in parentheses.}}
\end{table}

\begin{table}[ht]
\setlength{\tabcolsep}{4pt}
        \caption{Patent value and science quality (journal impact factor of SNPL reference) \label{reg:pat:sci:imp}}
        \centering
        \small
 \renewcommand{\arraystretch}{1.1}  
        \begin{tabular}{@{}l*{6}{S}@{}} \toprule
                    &\multicolumn{1}{c}{(1)}&\multicolumn{1}{c}{(2)}&\multicolumn{1}{c}{(3)}&\multicolumn{1}{c}{(4)}&\multicolumn{1}{c}{(5)}&\multicolumn{1}{c}{(6)}\\
DV (log):           & {5y Cit US}& {5y Cit EP}&{US Claim Length}&{EP Claim Length}&{USD Values}&{EUR Values}\\
\midrule
JIF SNPL ref (max)  &       0.117&       0.087&      -0.035&      -0.142&      -0.007&       0.246\\
                    &     (68.24)&     (67.56)&    (-26.79)&    (-31.82)&     (-0.71)&      (3.48)\\
\midrule
Patent-level controls&    {{Base}}&    {{Base}}&    {{Base}}&    {{Base}}&    {{Base}}&    {{Base}}\\
Patent applicant FE &      {{No}}&      {{No}}&      {{No}}&      {{No}}&      {{No}}&      {{No}}\\
Adj.\ R-Square      &     {0.149}&     {0.064}&     {0.157}&     {0.328}&     {0.112}&     {0.045}\\
Observations        &   {3929139}&   {3929139}&   {2289880}&   {1106701}&    {773969}&     {10261}\\
\bottomrule
\end{tabular}
\begin{tabular}{@{}l*{8}{S}} & & \\ \toprule &\multicolumn{1}{c}{{(7)}} &\multicolumn{1}{c}{{(8)}} &\multicolumn{1}{c}{{(9)}} &\multicolumn{1}{c}{{(10)}} &\multicolumn{1}{c}{{(11)}}&\multicolumn{1}{c}{{(12)}}\\
DV (log):           & {5y Cit US}& {5y Cit EP}&{US Claim Length}&{EP Claim Length}&{USD Values}&{EUR Values}\\
\midrule
JIF SNPL ref (max)  &       0.028&       0.056&      -0.039&      -0.102&      -0.056&       0.210\\
                    &     (14.94)&     (37.17)&    (-25.76)&    (-18.34)&     (-5.08)&      (2.99)\\
\midrule
Patent-level controls&     {{All}}&     {{All}}&     {{All}}&     {{All}}&     {{All}}&     {{All}}\\
Patent applicant FE &      {{No}}&      {{No}}&      {{No}}&      {{No}}&      {{No}}&      {{No}}\\
Adj.\ R-Square      &     {0.255}&     {0.097}&     {0.160}&     {0.329}&     {0.124}&     {0.065}\\
Observations        &   {3929139}&   {3929139}&   {2289880}&   {1106701}&    {773969}&     {10261}\\
\bottomrule
\end{tabular}
\begin{tabular}{@{}l*{8}{S}} & & \\ \toprule &\multicolumn{1}{c}{{(13)}} &\multicolumn{1}{c}{{(14)}} &\multicolumn{1}{c}{{(15)}} &\multicolumn{1}{c}{{(16)}} &\multicolumn{1}{c}{{(17)}}&\multicolumn{1}{c}{{(18)}}\\
DV (log):           & {5y Cit US}& {5y Cit EP}&{US Claim Length}&{EP Claim Length}&{USD Values}&{EUR Values}\\
\midrule
JIF SNPL ref (max)  &       0.015&       0.044&      -0.031&      -0.075&       0.020&       0.203\\
                    &      (6.83)&     (24.41)&    (-15.85)&    (-12.30)&      (4.18)&      (1.49)\\
\midrule
Patent-level controls&     {{All}}&     {{All}}&     {{All}}&     {{All}}&     {{All}}&     {{All}}\\
Patent applicant FE &     {{Yes}}&     {{Yes}}&     {{Yes}}&     {{Yes}}&     {{Yes}}&     {{Yes}}\\
Adj.\ R-Square      &     {0.359}&     {0.170}&     {0.260}&     {0.394}&     {0.891}&     {0.116}\\
Observations        &   {3385640}&   {3385640}&   {1931070}&    {993200}&    {734425}&      {5356}\\
\bottomrule
\end{tabular}

      \legend{\footnotesize{\textbf{Notes:} All reported values are elasticities.
      \textit{3y Cit SNPL ref (max)} is a measure of SNPL science quality corresponding to the maximum JIF across scientific publications appearing as SNPL references in a patent. Patent-level controls ``Base'' include technology fields and first filing year pair FEs. Patent-level controls ``All'' further include FEs for SNPL reference counts, patent reference counts and number of inventors. Patent applicant  FEs are derived from the first applicant on the grant publication. Robust standard errors. T-statistics in parentheses.}}
\end{table}

\begin{landscape}

\begin{table}[ht]
\setlength{\tabcolsep}{8pt}
        \caption{Patent value and science quality (alternative science quality indicators) \label{reg:pat:sci:cit:alt}}
        \centering
 \renewcommand{\arraystretch}{1.2}  
        \begin{tabular}{@{}l*{8}{S}@{}} \toprule
                    &\multicolumn{1}{c}{(1)}&\multicolumn{1}{c}{(2)}&\multicolumn{1}{c}{(3)}&\multicolumn{1}{c}{(4)}&\multicolumn{1}{c}{(5)}&\multicolumn{1}{c}{(6)}&\multicolumn{1}{c}{(7)}&\multicolumn{1}{c}{(8)}\\
DV (log):           & {5y Cit US}& {5y Cit US}& {5y Cit US}& {5y Cit US}& {5y Cit US}& {5y Cit US}& {5y Cit US}& {5y Cit US}\\
\midrule
SNPL science quality indicators: \\ 
\addlinespace
3y Cit SNPL ref (max)&       0.037&       0.027&            &            &            &            &            &            \\
                    &     (45.46)&     (29.68)&            &            &            &            &            &            \\
\addlinespace
3y Cit SNPL ref (sum)&            &            &       0.037&       0.026&            &            &            &            \\
                    &            &            &     (46.04)&     (29.31)&            &            &            &            \\
\addlinespace
3y Cit SNPL ref (avg)&            &            &            &            &       0.042&       0.030&            &            \\
                    &            &            &            &            &     (47.04)&     (29.82)&            &            \\
\addlinespace
3y Cit SNPL ref (sq)&            &            &            &            &            &            &       0.038&       0.027\\
                    &            &            &            &            &            &            &     (46.24)&     (29.91)\\
\midrule
Patent-level controls&     {{All}}&     {{All}}&     {{All}}&     {{All}}&     {{All}}&     {{All}}&     {{All}}&     {{All}}\\
Patent applicant FE &      {{No}}&     {{Yes}}&      {{No}}&     {{Yes}}&      {{No}}&     {{Yes}}&      {{No}}&     {{Yes}}\\
Adj.\ R-Square      &     {0.262}&     {0.362}&     {0.262}&     {0.362}&     {0.262}&     {0.362}&     {0.262}&     {0.362}\\
Observations        &   {4319309}&   {3763831}&   {4319309}&   {3763831}&   {4319309}&   {3763831}&   {4319309}&   {3763831}\\
\bottomrule
\end{tabular}

      \legend{\footnotesize{\textbf{Notes:} All reported values are elasticities.
      The table present results for alternative criteria of aggregation at the patent level of the science quality of SNPL references. The dependent variable is the 5-year count of patent forward-citations by US patents. Patent-level controls ``All'' include technology fields and first filing year pair FEs, FEs for SNPL reference counts, patent reference counts and number of inventors. Patent applicant  FEs are derived from the first applicant on the grant publication. Robust standard errors. T-statistics in parentheses.}}
\end{table}
\end{landscape}

\begin{table}[ht]
\setlength{\tabcolsep}{10pt}
        \caption{Patent value and science quality (self-references) \label{reg:pat:sci:cit:self}}
        \centering
 \renewcommand{\arraystretch}{1.2}  
        \begin{tabular}{@{}l*{4}{S}@{}} \toprule
Self-references                    &\multicolumn{2}{c}{{Excluded}}&\multicolumn{2}{c}{{Only}}\\
                    &\multicolumn{1}{c}{(1)}&\multicolumn{1}{c}{(2)}&\multicolumn{1}{c}{(3)}&\multicolumn{1}{c}{(4)}\\
DV (log):           & {5y Cit US}& {5y Cit US}& {5y Cit US}& {5y Cit US}\\
\midrule
3y Cit SNPL ref (max)&       0.033&       0.024&       0.040&       0.028\\
                    &     (36.00)&     (23.58)&     (24.21)&     (14.38)\\
\midrule
Patent-level controls&     {{All}}&     {{All}}&     {{All}}&     {{All}}\\
Patent applicant FE &      {{No}}&     {{Yes}}&      {{No}}&     {{Yes}}\\
Adj.\ R-Square      &     {0.255}&     {0.358}&     {0.235}&     {0.339}\\
Observations        &   {4100590}&   {3555570}&   {3684125}&   {3152984}\\
\bottomrule
\end{tabular}
\begin{tabular}{@{}l*{4}{S}}  && \\ \toprule &\multicolumn{1}{c}{{(5)}} &\multicolumn{1}{c}{{(6)}} &\multicolumn{1}{c}{{(7)}} &\multicolumn{1}{c}{{(8)}} \\ 
DV (log):           & {5y Cit EP}& {5y Cit EP}& {5y Cit EP}& {5y Cit EP}\\
\midrule
3y Cit SNPL ref (max)&       0.025&       0.018&       0.046&       0.037\\
                    &     (35.20)&     (22.53)&     (34.07)&     (23.16)\\
\midrule
Patent-level controls&     {{All}}&     {{All}}&     {{All}}&     {{All}}\\
Patent applicant FE &      {{No}}&     {{Yes}}&      {{No}}&     {{Yes}}\\
Adj.\ R-Square      &     {0.090}&     {0.160}&     {0.084}&     {0.151}\\
Observations        &   {4100590}&   {3555570}&   {3684125}&   {3152984}\\
\bottomrule
\end{tabular}

      \legend{\footnotesize{\textbf{Notes:} All reported values are elasticities.
      \textit{3y Cit SNPL ref (max)} is a measure of SNPL science quality corresponding to the maximum 3-year citation count across scientific publications appearing as SNPL references in a patent. 
      Patent-level controls ``All'' further include FEs for SNPL reference counts, patent reference counts and number of inventors. Patent applicant  FEs are derived from the first applicant on the grant publication. Robust standard errors. T-statistics in parentheses.}}
\end{table}

\begin{table}[ht]
\setlength{\tabcolsep}{8pt}
        \caption{Patent value and science quality (by technology area of patent) \label{reg:pat:sci:cit:area}}
        \centering
 \renewcommand{\arraystretch}{1.2}  
        \begin{tabular}{@{}l*{4}{S}@{}} \toprule
Technology area                    &\multicolumn{1}{c}{{Electrical Eng}}&\multicolumn{1}{c}{{Instruments}}&\multicolumn{1}{c}{{Chemistry}}&\multicolumn{1}{c}{{Mechanical Eng}}\\
                    &\multicolumn{1}{c}{(1)}&\multicolumn{1}{c}{(2)}&\multicolumn{1}{c}{(3)}&\multicolumn{1}{c}{(4)}\\
DV (log):           & {5y Cit US}& {5y Cit US}& {5y Cit US}& {5y Cit US}\\
\midrule
3y Cit SNPL ref (max)&       0.028&       0.030&       0.060&       0.048\\
                    &     (21.60)&     (15.51)&     (45.56)&     (11.32)\\
\addlinespace
Constant            &       1.645&       1.471&       1.047&       1.080\\
                    &   (1754.24)&    (979.79)&    (518.65)&   (1323.25)\\
\midrule
Patent-level controls&     {{All}}&     {{All}}&     {{All}}&     {{All}}\\
Patent applicant FE &      {{No}}&      {{No}}&      {{No}}&      {{No}}\\
Adj.\ R-Square      &     {0.204}&     {0.232}&     {0.235}&     {0.176}\\
Observations        &   {1542153}&    {713973}&    {779607}&    {953557}\\
\bottomrule
\end{tabular}
\begin{tabular}{@{}l*{4}{S}} \phantom{Technology area} & {\phantom{Electrical Eng}} & {\phantom{Instruments}} & {\phantom{Chemistry}} & {\phantom{Mechanical Eng}} \\ \toprule &\multicolumn{1}{c}{{(5)}} &\multicolumn{1}{c}{{(6)}} &\multicolumn{1}{c}{{(7)}} &\multicolumn{1}{c}{{(8)}} \\ 
DV (log):           & {5y Cit US}& {5y Cit US}& {5y Cit US}& {5y Cit US}\\
\midrule
3y Cit SNPL ref (max)&       0.026&       0.024&       0.040&       0.032\\
                    &     (17.75)&     (10.43)&     (26.42)&      (6.28)\\
\midrule
Patent-level controls&     {{All}}&     {{All}}&     {{All}}&     {{All}}\\
Patent applicant FE &     {{Yes}}&     {{Yes}}&     {{Yes}}&     {{Yes}}\\
Adj.\ R-Square      &     {0.316}&     {0.359}&     {0.324}&     {0.309}\\
Observations        &   {1359841}&    {570768}&    {652006}&    {755229}\\
\bottomrule
\end{tabular}

      \legend{\footnotesize{\textbf{Notes:} All reported values are elasticities.
      \textit{3y Cit SNPL ref (max)} is a measure of SNPL science quality corresponding to the maximum 3-year citation count across scientific publications appearing as SNPL references in a patent. 
      Patent-level controls ``All'' further include FEs for SNPL reference counts, patent reference counts and number of inventors. Patent applicant  FEs are derived from the first applicant on the grant publication. Robust standard errors. T-statistics in parentheses.}}
\end{table}

\begin{table}[ht]
\setlength{\tabcolsep}{8pt}
        \caption{Patent value and science quality (by patent applicant country) \label{reg:pat:sci:cit:ctry}}
        \centering
 \renewcommand{\arraystretch}{1.2}  
        \begin{tabular}{@{}l*{5}{S}@{}} \toprule
Patent applicant country                    &\multicolumn{1}{c}{{China}}&\multicolumn{1}{c}{{Europe}}&\multicolumn{1}{c}{{Japan}}&\multicolumn{1}{c}{{South Korea}}&\multicolumn{1}{c}{{United States}}\\
                    &\multicolumn{1}{c}{(1)}&\multicolumn{1}{c}{(2)}&\multicolumn{1}{c}{(3)}&\multicolumn{1}{c}{(4)}&\multicolumn{1}{c}{(5)}\\
DV (log):           & {5y Cit US}& {5y Cit US}& {5y Cit US}& {5y Cit US}& {5y Cit US}\\
\midrule
3y Cit SNPL ref (max)&       0.020&       0.042&       0.036&       0.014&       0.032\\
                    &      (3.14)&     (25.00)&     (18.40)&      (3.14)&     (29.20)\\
\midrule
Patent-level controls&     {{All}}&     {{All}}&     {{All}}&     {{All}}&     {{All}}\\
Patent applicant FE &      {{No}}&      {{No}}&      {{No}}&      {{No}}&      {{No}}\\
Adj.\ R-Square      &     {0.218}&     {0.287}&     {0.171}&     {0.176}&     {0.268}\\
Observations        &     {54062}&    {948243}&    {869068}&    {149389}&   {2032330}\\
\bottomrule
\end{tabular}
\begin{tabular}{@{}l*{5}{S}}  \phantom{Patent applicant country} & {\phantom{China}} & {\phantom{Europe}} & {\phantom{Japan}} & {\phantom{South Korea}}  & {\phantom{United States}} \\ \toprule &\multicolumn{1}{c}{{(6)}} &\multicolumn{1}{c}{{(7)}} &\multicolumn{1}{c}{{(8)}} &\multicolumn{1}{c}{{(9)}} &\multicolumn{1}{c}{{(10)}} \\ 
DV (log):           & {5y Cit EP}& {5y Cit EP}& {5y Cit EP}& {5y Cit EP}& {5y Cit EP}\\
\midrule
3y Cit SNPL ref (max)&       0.004&       0.033&       0.021&       0.017&       0.033\\
                    &      (0.86)&     (23.78)&     (13.77)&      (4.99)&     (38.63)\\
\midrule
Patent-level controls&     {{All}}&     {{All}}&     {{All}}&     {{All}}&     {{All}}\\
Patent applicant FE &      {{No}}&      {{No}}&      {{No}}&      {{No}}&      {{No}}\\
Adj.\ R-Square      &     {0.126}&     {0.090}&     {0.073}&     {0.118}&     {0.133}\\
Observations        &     {54062}&    {948243}&    {869068}&    {149389}&   {2032330}\\
\bottomrule
\end{tabular}

      \legend{\footnotesize{\textbf{Notes:} All reported values are elasticities.
      \textit{3y Cit SNPL ref (max)} is a measure of SNPL science quality corresponding to the maximum 3-year citation count across scientific publications appearing as SNPL references in a patent. 
      Patent-level controls ``All'' further include FEs for SNPL reference counts, patent reference counts and number of inventors. Patent applicant  FEs are derived from the first applicant on the grant publication. Robust standard errors. T-statistics in parentheses.}}
\end{table}

\clearpage

\begin{table}[ht]
\setlength{\tabcolsep}{10pt}
        \caption{Patent value and science quality (interdisciplinarity) \label{reg:pat:sci:cit:inter}}
        \centering
 \renewcommand{\arraystretch}{1.2}  
        \begin{tabular}{@{}l*{4}{S}@{}} \toprule
                    &\multicolumn{1}{c}{(1)}&\multicolumn{1}{c}{(2)}&\multicolumn{1}{c}{(3)}&\multicolumn{1}{c}{(4)}\\
DV (log):           & {5y Cit US}& {5y Cit US}& {5y Cit EP}& {5y Cit EP}\\
\midrule
Interdisciplinary   &       0.054&       0.047&       0.040&       0.037\\
                    &     (15.47)&     (12.41)&     (15.35)&     (12.73)\\
\addlinespace
3y Cit SNPL ref (max) $\times$ Single Discipline&       0.042&       0.031&       0.035&       0.027\\
                    &     (45.13)&     (29.78)&     (47.66)&     (32.32)\\
\addlinespace
3y Cit SNPL ref (max) $\times$ Interdisciplinary&       0.032&       0.024&       0.023&       0.017\\
                    &     (30.66)&     (20.53)&     (27.80)&     (18.68)\\
\midrule
Patent-level controls&     {{All}}&     {{All}}&     {{All}}&     {{All}}\\
Patent applicant FE &      {{No}}&     {{Yes}}&      {{No}}&     {{Yes}}\\
Adj.\ R-Square      &     {0.262}&     {0.362}&     {0.100}&     {0.169}\\
Observations        &   {4319309}&   {3763831}&   {4319309}&   {3763831}\\
\bottomrule
\end{tabular}

      \legend{\footnotesize{\textbf{Notes:} All reported values are elasticities.
      \textit{3y Cit SNPL ref (max)} is a measure of SNPL science quality corresponding to the maximum 3-year citation count across scientific publications appearing as SNPL references in a patent. The interdisciplinarity status is taken from the most cited scientific publication appearing as SNPL reference in a patent. 
      Patent-level controls ``All'' further include FEs for SNPL reference counts, patent reference counts and number of inventors. Patent applicant  FEs are derived from the first applicant on the grant publication. Robust standard errors. T-statistics in parentheses.}}
\end{table}

\clearpage

\begin{table}[ht]
\setlength{\tabcolsep}{8pt}
        \caption{Patent value and scientific impact (US citations) (by frontier distance) \label{reg:pat:sci:cit:frontier}}
        \centering
 \renewcommand{\arraystretch}{1.2}  
        \begin{tabular}{@{}l*{4}{S}@{}} \toprule & \multicolumn{1}{c}{(1)} & &\multicolumn{1}{c}{(2)} \\ DV (log): & \multicolumn{1}{c}{5y Cit US} & &\multicolumn{1}{c}{5y Cit US} \\
\midrule
Distance to frontier: \\ 
0                   &       0.818&    (416.50)&       0.753&    (327.87)\\
1                   &       0.595&    (327.21)&       0.567&    (287.88)\\
2                   &       0.363&    (191.49)&       0.403&    (188.36)\\
3                   &       0.232&    (101.58)&       0.283&     (99.75)\\
4                   &       0.150&     (48.38)&       0.212&     (51.28)\\
5                   &       0.144&     (38.10)&       0.215&     (39.75)\\
6                   &       0.157&     (43.21)&       0.257&     (45.98)\\
7                   &       0.149&     (46.75)&       0.270&     (54.86)\\
8                   &       0.121&     (41.76)&       0.249&     (56.85)\\
9                   &       0.080&     (27.94)&       0.206&     (48.36)\\
10                  &       0.035&     (11.37)&       0.166&     (36.66)\\
3y Cit SNPL ref (max)&       0.056&    (194.53)&            &            \\
3y Cit SNPL ref (max) $\times$ 0&            &            &       0.089&    (144.39)\\
3y Cit SNPL ref (max) $\times$ 1&            &            &       0.070&    (147.30)\\
3y Cit SNPL ref (max) $\times$ 2&            &            &       0.037&     (62.50)\\
3y Cit SNPL ref (max) $\times$ 3&            &            &       0.027&     (25.62)\\
3y Cit SNPL ref (max) $\times$ 4&            &            &       0.018&     (10.14)\\
3y Cit SNPL ref (max) $\times$ 5&            &            &       0.018&      (8.19)\\
3y Cit SNPL ref (max) $\times$ 6&            &            &       0.007&      (3.50)\\
3y Cit SNPL ref (max) $\times$ 7&            &            &      -0.002&     (-0.90)\\
3y Cit SNPL ref (max) $\times$ 8&            &            &      -0.002&     (-1.62)\\
3y Cit SNPL ref (max) $\times$ 9&            &            &       0.000&      (0.35)\\
3y Cit SNPL ref (max) $\times$ 10&            &            &      -0.001&     (-0.83)\\
\midrule
Patent-level controls&    {{Base}}&            &    {{Base}}&            \\
Patent applicant FE &      {{No}}&            &      {{No}}&            \\
Adj.\ R-Square      &     {0.187}&            &     {0.189}&            \\
Observations        &   {4378579}&            &   {4378579}&            \\
\bottomrule
\end{tabular}

      \legend{\footnotesize{\textbf{Notes:} All reported values are elasticities.
      Patent-level controls ``Base'' include technology fields and first filing year pair FEs.  \textit{3y Cit SNPL ref (max)} is a measure of SNPL science quality corresponding to the maximum 3-year citation count across scientific publications appearing as SNPL references in a patent. Robust standard errors. T-statistics in parentheses.
      }}
\end{table}

\clearpage

\begin{table}[ht]
\setlength{\tabcolsep}{10pt}
        \caption{Patent value and scientific impact (by time-distance) \label{reg:pat:sci:cit:lag}}
        \centering
 \renewcommand{\arraystretch}{1.2}  
        \begin{tabular}{@{}l*{4}{S}@{}} \toprule
                    &\multicolumn{1}{c}{(1)}&\multicolumn{1}{c}{(2)}&\multicolumn{1}{c}{(3)}&\multicolumn{1}{c}{(4)}\\
DV (log):           & {5y Cit US}& {5y Cit US}& {5y Cit EP}& {5y Cit EP}\\
\midrule
SNPL time distance: \\ 
\addlinespace
Short               &       0.141&       0.116&       0.089&       0.076\\
                    &     (33.37)&     (25.58)&     (28.11)&     (21.56)\\
\addlinespace
Medium              &       0.052&       0.040&       0.039&       0.028\\
                    &     (12.09)&      (8.56)&     (12.14)&      (7.79)\\
\addlinespace
3y Cit SNPL ref (max) $\times$ Short&       0.039&       0.031&       0.033&       0.027\\
                    &     (36.56)&     (26.38)&     (38.86)&     (28.03)\\
\addlinespace
3y Cit SNPL ref (max) $\times$ Medium&       0.037&       0.027&       0.027&       0.021\\
                    &     (33.39)&     (21.97)&     (30.41)&     (20.90)\\
\addlinespace
3y Cit SNPL ref (max) $\times$ Long&       0.031&       0.019&       0.027&       0.018\\
                    &     (26.87)&     (14.92)&     (30.00)&     (17.35)\\
\midrule
Patent-level controls&     {{All}}&     {{All}}&     {{All}}&     {{All}}\\
Patent applicant FE &      {{No}}&     {{Yes}}&      {{No}}&     {{Yes}}\\
Adj.\ R-Square      &     {0.263}&     {0.363}&     {0.101}&     {0.169}\\
Observations        &   {4319309}&   {3763831}&   {4319309}&   {3763831}\\
\bottomrule
\end{tabular}

      \legend{\footnotesize{\textbf{Notes:} All reported values are elasticities.
      \textit{3y Cit SNPL ref (max)} is a measure of SNPL science quality corresponding to the maximum 3-year citation count across scientific publications appearing as SNPL references in a patent. \textit{Short}, \textit{Medium} and \textit{long} time-distance are dummies for the tertiles of time-distance. 
      Patent-level controls ``All'' further include FEs for SNPL reference counts, patent reference counts and number of inventors. Patent applicant  FEs are derived from the first applicant on the grant publication. Robust standard errors. T-statistics in parentheses.}}    
\end{table}

\clearpage

\singlespacing

\bibliographystyle{chicago}
\bibliography{bibliography_npl}

\end{document}